\documentclass[%
reprint,superscriptaddress, amsmath,amssymb,aps,prx, 
floatfix,]{revtex4-2}
\usepackage{graphicx}
\usepackage{dcolumn}
\usepackage{bm}
\usepackage{hyperref}
\hypersetup{colorlinks=true, citecolor=blue, urlcolor=blue, linkcolor=blue}
\usepackage{braket}
\usepackage{accents}
\usepackage{multirow,microtype}
\usepackage{dcolumn,booktabs}
\usepackage[version=4]{mhchem}

\begin{document}

\preprint{APS/123-QED}

\title{Unified Differentiable Learning of Electric Response}

\author{Stefano Falletta}
\email{stefanofalletta@g.harvard.edu}
\affiliation{John A.\ Paulson School of Engineering and Applied Sciences, Harvard University, Cambridge, MA, USA}

\author{Andrea Cepellotti}
\affiliation{John A.\ Paulson School of Engineering and Applied Sciences, Harvard University, Cambridge, MA, USA}

\author{Anders Johansson}
\affiliation{John A.\ Paulson School of Engineering and Applied Sciences, Harvard University, Cambridge, MA, USA}

\author{Chuin Wei Tan}
\affiliation{John A.\ Paulson School of Engineering and Applied Sciences, Harvard University, Cambridge, MA, USA}

\author{Albert Musaelian}
\affiliation{John A.\ Paulson School of Engineering and Applied Sciences, Harvard University, Cambridge, MA, USA}

\author{Cameron J. Owen}
\affiliation{Department of Chemistry and Chemical Biology, Harvard University, Cambridge, MA, USA}

\author{Boris Kozinsky}
\email{bkoz@g.harvard.edu}
\affiliation{John A.\ Paulson School of Engineering and Applied Sciences, Harvard University, Cambridge, MA, USA}
\affiliation{Robert Bosch LLC Research and Technology Center, Watertown, MA, USA}

\date{\today}

\begin{abstract} 
Predicting response of materials to external stimuli is a primary objective of computational materials science. However, current methods are limited to small-scale simulations due to the unfavorable scaling of computational costs. Here, we implement an equivariant machine-learning framework where response properties stem from exact differential relationships between a generalized potential function and applied external fields. Focusing on responses to electric fields, the method predicts electric enthalpy, forces, polarization, Born charges, and polarizability within a unified model enforcing the full set of exact physical constraints, symmetries and conservation laws. Through application to $\alpha$-\ce{SiO2}, we demonstrate that our approach can be used for predicting vibrational and dielectric properties of materials, and for conducting large-scale dynamics under arbitrary electric fields at unprecedented accuracy and scale. We apply our method to ferroelectric \ce{BaTiO3} and capture the temperature-dependence and time evolution of hysteresis, revealing the underlying microscopic mechanisms of nucleation and growth that govern ferroelectric domain switching.
\end{abstract}

\maketitle

The goal of computational materials science is to accurately predict experimentally measurable properties of real materials from first principles. Linear, nonlinear, and coupled responses to external stimuli define the functional properties of a wide class of materials including dielectrics, ferroelectrics, multiferroics, and piezoelectrics. 
Developing computational methods to calculate materials response to external stimuli has been a long-standing goal of first-principles electronic structure methods based on density functional theory (DFT). Perturbative or finite difference DFT approaches to response \cite{baroni2001RMP,gonze1997PRB_2} are limited to very small systems, due to the unfavorable scaling of computational costs. In recent years, machine-learning (ML) force fields have closed the gap between the accuracy of DFT calculations and the efficiency required for large-scale calculations, such as molecular dynamics (MD) simulations, determination of elastic constants, and phonon spectra. The accuracy of ML force fields has been significantly enhanced by incorporating exact $O(3)$ symmetry group equivariance, starting with the NequIP model and subsequent approaches  \cite{batzner2023NC,musaelian2023NC,batatia2022arxiv,nigam2022JCP,geiger2022e3nn} to learn the potential energy as a function of atomic coordinates. In this context, there is a need for a ML framework for generalized potentials, which can depend in a nonlinear and coupled way on a number of parameters such as state variables or external fields.

Among various types of perturbations, responses of materials to external electric fields are some of the most important. Vibrational and dielectric responses of crystalline, disordered and liquid materials can be determined from the dynamics of polarization and polarizability \cite{neumann1983MP,neumann1983CPL}. Performance of ferroelectric devices, such as non-volatile memories, sensors, and actuators, is governed by hysteresis, polarization switching, and domain wall motion \cite{shin2007nature,liu2016nature}, whose microscopic mechanisms are not yet fully understood due to the limitations of both first principle computations and experimental measurements. Quantitatively accurate simulations that can reveal the microscopic mechanisms of switching dynamics must simultaneously account for the presence of external electric fields and handle the complexity and large scales to capture the influence of defects and surfaces on the nucleation and growth of ferroelectric domains.

Driven by the need of understanding the dielectric and ferroelectric properties of materials from first principles \cite{resta1994RMP}, the modern theory of polarization \cite{resta1994RMP,kingsmith1993PRB,spaldin2012JSSC} and electric enthalpy functionals \cite{nunes2001PRB,umari2002PRL} have been introduced. However, the significant computational cost of DFT limits the ability to simulate realistic materials and devices. To address this limitation, ML approaches have been proposed to predict polarization \cite{christensen2019JCP,veit2020JCP,krishnamoorthy2021PRL,gastegger2021CS,staackeMLST2022,gigli2022npj,schutt2021ICML,schienbein2023JCTC,shao2023NC,zhang2022JCP,joll2024arxiv,shimizu2023STAM,choudhary2020npj,fang2024arxiv,schmiedmayer2024arxiv} and other response properties, such as Born charges \cite{choudhary2020npj,shimizu2023STAM,zhang2022JCP,joll2024arxiv,fang2024arxiv,schmiedmayer2024arxiv}, polarizability \cite{wilkins2019PNAS,sommers2020PCCP,zhang2022JCP,fang2024arxiv}, dielectric constants \cite{choudhary2020npj,takahashi2020PRM,krishnamoorthy2021PRL,gigli2022npj,zhang2022JCP,joll2024arxiv}, infrared spectra \cite{choudhary2020npj,gastegger2021CS,schienbein2023JCTC,shao2023NC,kapil2023FD,zhang2022JCP,joll2024arxiv,schmiedmayer2024arxiv}, and Raman spectra \cite{sommers2020PCCP,gastegger2021CS,kapil2023FD,berger2024PRM,fang2024arxiv}, with applications to molecules \cite{wilkins2019PNAS,gastegger2021CS,staackeMLST2022,shao2023NC,veit2020JCP,christensen2019JCP,fang2024arxiv}, liquid water \cite{choudhary2020npj,sommers2020PCCP,krishnamoorthy2021PRL,schienbein2023JCTC,kapil2023FD,zhang2022JCP,joll2024arxiv}, and solids \cite{gigli2022npj,shimizu2023STAM,kapil2023FD}. 
Most of these ML methods are formulated in a disjoint manner, with a conventional ML model trained to evolve atomic positions and a separate ML model trained to predict dielectric properties. However, this does not guarantee the enforcement of physical symmetries and conservation laws involving electric enthalpy, polarization, and Born charges. Approaches in which dielectric properties  are determined together with energy and forces within a single model \cite{christensen2019JCP,gastegger2021CS,shao2023NC,fang2024arxiv} have not been formulated for extended systems, where periodic boundaries and the multivalued nature of polarization pose difficulties \cite{schmiedmayer2024arxiv}.

We introduce a unified differential framework for learning the generalized potential energy and the response functions to external stimuli within a single ML model. This is achieved by determining the response functions as derivatives of the generalized potential energy with respect to atomic coordinates and perturbation parameters. Because our method is based on exact differential relations between the generalized potential energy and the observable response quantities, it enforces both physical symmetries and conservation laws involving physical quantities, which cannot be achieved when using separate ML models for each response function. We illustrate this approach in the case of an applied electric field. Specifically, we learn the electric enthalpy as a function of atomic positions and electric field, and derive polarization by differentiating the electric enthalpy with respect to the electric field, the Born charges by differentiating the polarization with respect to the atomic positions, and the polarizability by differentiating the polarization with respect to the electric field. This formalism guarantees momentum conservation, the acoustic sum rule for Born charges, the polarization being a conservative vector field, and the electric enthalpy conservation in ML molecular dynamics (MLMD) and in cyclic adiabatic evolutions involving changes of the electric field. Our architecture augments the inputs with parameters describing the system perturbation, such that model differentiation with respect to these parameters allows the model to train on additional physical quantities. This approach differs from that of physically informed neural networks \cite{raissi2017arxiv}, in which the loss function incorporates additional regularization terms pertaining to differential expressions involving only the output of the model. We validate our method by calculating infrared spectrum, frequency-dependent dielectric constant and screening effects in $\alpha$-\ce{SiO2}, finding excellent agreement with results from density-functional perturbation theory. We further demonstrate the capability of our approach to perform ML-driven molecular dynamics in the presence of an electric field and apply our method to study the temperature-dependent ferroelectric properties of tetragonal \ce{BaTiO3}. Specifically, we calculate the ferroelectric hysteresis and investigate the underlying dipole dynamics, providing real-time description of nucleation and evolution mechanisms of ferroelectric domains. In this way, our work paves the way to efficient and accurate large-scale studies of dielectric and ferroelectric properties of crystalline, disordered and liquid materials, far beyond the reach of standard quantum mechanical methods in time and length scales.

\begin{figure*}[t!]
    \centerline{\includegraphics[width=0.95\linewidth]{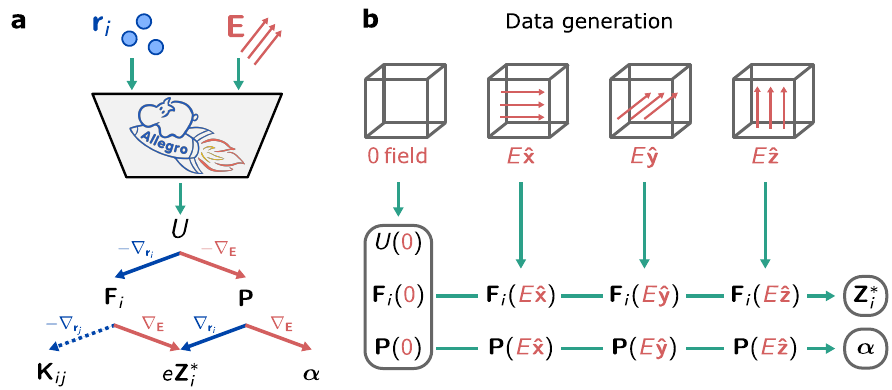}}
    \caption{\textbf{Unified ML formulation and framework for data generation.} \textbf{a}, Schematics outlining our neural network formulation that simultaneously learns to predict electric enthalpy $U$, forces $\mathbf F_{i}$, polarization $\mathbf P$, Born charges $\mathbf Z^*_{i}$, and polarizability $\bm\alpha$ given in input the atomic positions $\mathbf r_{i}$ and the electric field $\mathbf E$. The force constant $\mathbf K_{ij}$ is shown for completeness, but is not used for demonstrating our approach in this work. Here, $i$ denotes the atom index. \textbf{b}, Workflow for generating DFT data through calculations with finite electric fields for a given set of frames. The DFT data used for training and validation of the ML model are calculated in the limit of zero electric field, and are circled for ease of reference.}
    \label{fig:model}
\end{figure*}

\vspace{0.3cm} \noindent {\large \textbf{Results}} \vspace{0.1cm}\\
\textbf{Machine-learning framework for response} \\ 
We focus on learning the generalized potential energy $U$ of a system that depends on atomic coordinates and on a set of parameters, which include external fields. Following a Taylor expansion expression [cf.\ equation \eqref{eq:e_expansion} in Methods], we introduce a framework for learning the generalized potential energy and response functions within a single unified ML model. This is based on the idea that differentiating the generalized potential energy with respect to its variables automatically yields response properties for each atomic configuration. The training of the model is achieved by minimizing a loss function with contributions pertaining to each response property. This framework is related to the concept of Sobolev training \cite{czarnecki2017arxiv} in that each loss term consists of differences between training label and corresponding gradient of that energy. In the context of response theory, we achieve generality and ability to train the model to describe a system's response to variation of any parameter, with other parameters held at arbitrary constant values. This framework is versatile and applicable to various ML methods, encompassing both invariant and equivariant neural networks \cite{batzner2023NC,musaelian2023NC}, along with kernel-based methods \cite{vandermause2020npj}. As response properties may exhibit nonlinear and coupled dependencies on more than one field, we employ neural networks for capturing intricate dependencies of the generalized potential energy on its inputs through back-propagation of the gradients. The translation invariance of the generalized potential energy combined with the exact derivative relations for response functions enforce exact physical symmetries and conservation laws, including momentum conservation, the electric enthalpy conservation during dynamics, and conservation laws involving the response functions. While the discussion in this work deals only with the microscopic generalized potential energy, we note that we have concurrently formulated a general differentiable formalism for learning temperature-dependent free energy models in the context of dimensionality reduction, such as molecular coarse graining \cite{duschatko2024arxiv}.

As a concrete example of this approach, we consider the generalized potential energy of a system subject to an electric field, namely the electric enthalpy \cite{nunes2001PRB,umari2002PRL}. We implement this model in Allegro \cite{musaelian2023NC}, which offers the accuracy and data efficiency advantages of an equivariant neural network combined with state-of-the-art scalability due to strict locality. As illustrated in Fig.\ \ref{fig:model}a, we include the electric field among the inputs of the model together with the atomic positions. Assuming the linear response regime, we focus on training the models on energies (electric enthalpies), forces, polarizations, and Born charges at zero electric field. This allows the implementation to be a simple modification of the conventional Allegro ML force field architecture. For a given set of uncorrelated atomic configurations, the training data is generated by performing DFT calculations both in the absence and in the presence of a small electric field along each of the three Cartesian directions. Then, finite-difference approximations are used to determine Born charges and polarizabilities (cf.\ Fig.\ \ref{fig:model}b). For training and validation, we use labels calculated in the limit of zero electric field.
Once a model is trained, it outputs the electric enthalpy and derives therefrom forces, polarization, Born charges, and polarizability by taking first and second derivatives of the output with respect to atomic positions and electric field, at zero electric field. The contributions to energy, forces, and polarization due to the presence of an electric field are included analytically within the linear response approximation (see Methods). Subsequently, large-scale structural relaxations and MLMD in the presence of an electric field can be performed with \textsc{lammps} \cite{thompson2022CPC} through a dedicated interface that we developed. Further details on the theory, neural network architecture, and simulations under external electric fields are given in Methods.

Our framework offers several physical advantages. First, it is physically elegant as it predicts electric enthalpy and response quantities within a single unified model through exact physical relations. Second, physical symmetries are enforced by construction. In this regard, electric enthalpy and polarization are invariant under translations, as they depend only on interatomic displacements within our ML model. Third, the physical symmetries and the exact constraints satisfied by our ML model enforce conservation laws. In particular, the momentum is conserved since the electric enthalpy is translation invariant and forces are calculated as gradients of the electric enthalpy with respect to atomic positions. Similarly, the acoustic sum rule for Born charges is satisfied, since polarization is translation invariant and Born charges are determined as gradients of polarization with respect to atomic positions. This reflects the charge neutrality of the system. Furthermore, the electric enthalpy is conserved in MLMD, since forces are calculated as gradients of the electric enthalpy with respect to atomic positions. Finally, polarization is guaranteed to be a conservative vector field, as it is calculated as gradient of the electric enthalpy with respect to the electric field. This implies the conservation of electric enthalpy in any cyclic adiabatic evolution involving changes in the electric field, which is relevant for studying response to oscillating fields. A more detailed discussion on physical symmetries and conservation laws, together with an extensive comparison with previous literature is provided in Methods.

\begin{figure}[t!]
    \centerline{\includegraphics[width=1.0\linewidth]{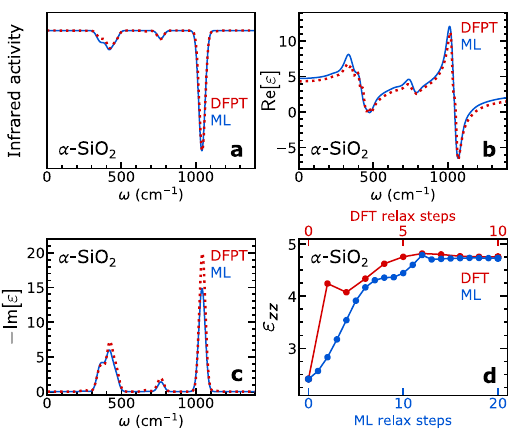}}
    \caption{\textbf{Dielectric properties of $\alpha$-\ce{SiO2}.} \textbf{a}, Infrared spectrum of $\alpha$-\ce{SiO2} at 300~K as a function of frequency $\omega$ obtained with MLMD and with DFPT. The amplitudes of the main infrared peak obtained in the two cases are matched as a guide to the eye.  \textbf{b}, Real part and \textbf{c}, imaginary part of the frequency-dependent dielectric constant of $\alpha$-\ce{SiO2} at 300~K as a function of frequency $\omega$ obtained with MLMD and with DFPT. Here, $\varepsilon = (\varepsilon_{11} + \varepsilon_{22} + \varepsilon_{33})/3$. In \textbf{a}-\textbf{c}, a Gaussian broadening of 20 cm$^{-1}$ is used. \textbf{d}, Dielectric constant $\varepsilon_{zz}$ of $\alpha$-\ce{SiO2} obtained through structural relaxations in the presence of an electric field of along $\mathbf z$ with our ML model and with DFT, starting from a pristine bulk structure.}
    \label{fig:results_SiO2}
\end{figure}

\vspace{0.3cm} \noindent \textbf{Vibrational and dielectric properties} \\
We begin by demonstrating our ML framework for investigating the vibrational and dielectric properties of $\alpha$-\ce{SiO2}. We train an ML model for $\alpha$-\ce{SiO2} using 200 frames of 72 atoms extracted from MD simulations employing a classical potential. Then, we construct a 24696-atom supercell and perform MLMD for 200 ps in the absence of electric field within the NVT ensemble. From the polarization dynamics, we determine the infrared spectrum of $\alpha$-\ce{SiO2}, identifying main peaks at frequencies of 1041, 420, and 765 cm$^{-1}$ (cf.\ Fig.\ \ref{fig:results_SiO2}a). Next, by analyzing the dynamics of polarization and polarizability, we determine the frequency-dependent dielectric constant, which we illustrate in Figs.\ \ref{fig:results_SiO2}b-c. For comparison, we calculate these quantities using density functional perturbation theory (DFPT) \cite{baroni2001RMP,gonze1997PRB_2} following the approach in Ref.\ \cite{giacomazzi2009PRB}. As illustrated in Figs.\ \ref{fig:results_SiO2}a-c, we find excellent agreement between the MLMD and DFPT results, thereby validating of our method. 

Next, we show that our formulation describes electronic and ionic screening effects in the presence of an electric field. Specifically, we focus on the determination of the static dielectric constant, which results from the difference between the polarization calculated in the presence of a small electric field for the structure relaxed with the electric field, and the polarization calculated in the absence of the electric field for the pristine bulk structure [see equation \eqref{eq:epsilon_P}]. This requires the ML model to accurately capture electric field contributions to electric enthalpy, polarization and forces [cf.\ equations \eqref{eq:enthalpy}, \eqref{eq:force_efield} and \eqref{eq:polarization_efield}], which is challenging as these contributions can be small and comparable to the accuracy of the model. We hence perform a structural relaxation under a finite electric field starting from a pristine bulk structure of $\alpha$-SiO$_2$ using our ML model. As illustrated in Fig.\ \ref{fig:results_SiO2}d, we find high-frequency and static dielectric constants equal to $\varepsilon_{zz}^\infty = 2.41$ and $\varepsilon_{zz}^0 = 4.75$, which essentially coincide with the respective DFT values. This demonstrates that our method successfully captures the electric-field contributions to the electronic structure, thereby further corroborating the validity of our formulation for performing dynamics under finite electric fields. Details on data generation and expressions used for infrared spectrum and dielectric constants are provided in Methods. 

\vspace{0.3cm} \noindent \textbf{Ferroelectric domain dynamics and hysteresis} \\ 
The computational study of ferroelectric properties from first principles, and especially dynamics of domain switching, is challenging due to the high computational cost involved. For instance, the analysis of the ferroelectric hysteresis requires performing multiple relaxations of sufficiently large structures under applied electric fields. In particular, determining the coercive field upon which the polarization switches sign requires progressively smaller changes in the electric field close to the transition, which becomes prohibitive for large systems to capture with DFT. In addition, performing such a task at finite temperature requires conducting MD simulations under electric fields, which is again intractable with DFT. 
Empirical bond-valence potential simulations combined with simple analytical models were used to study ferroelectric domain wall motion \cite{shin2007nature,liu2016nature}. 
Simulations with ML force fields with ad-hoc Born charges have also been performed \cite{akbarian2019PCCP,deguchi2023PSS}. However, such uncontrolled approximations in these effort have unpredictable accuracy, especially when quantifying switching dynamics in realistic devices in the presence of defects. In this regard, our method provides a scalable but rigorous first-principles-based comprehensive description of ferroelectric dynamics, with explicit accurate learning  of polarization and Born charges obtained from ab-initio calculations based on the modern theory of polarization. This is achieved by constructing a differential response theory model architecture based on state-of-the-art equivariant neural networks, offering a significant speedup in computation while maintaining quantum mechanical accuracy.

\begin{figure}[t!]
    \centerline{\includegraphics[width=1.0\linewidth]{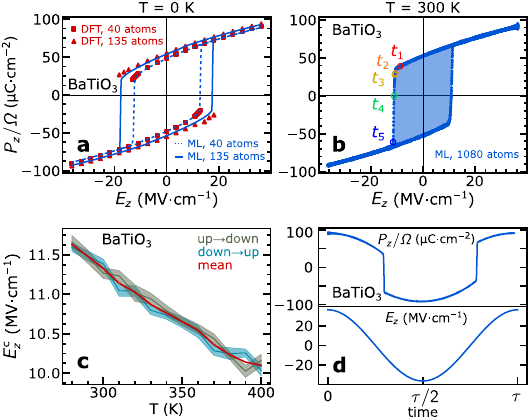}}
    \caption{\textbf{Ferroelectric properties of  \ce{BaTiO3}.} \textbf{a}, Ferroelectric hysteresis of polarization density $P_z/\Omega$ as a function of an applied electric field along $\mathbf z$ in \ce{BaTiO3} at 0~K, as obtained through structural relaxations with our ML model or with DFT. \textbf{b}, Ferroelectric hysteresis  of polarization density $P_z/\Omega$ as a function of an applied electric field along $\mathbf z$ in \ce{BaTiO3} at 300~K obtained with our ML model. Time signatures $t_i$ are used for illustrating the evolution of the polarized states. \textbf{c}, Coercive electric field $E^\text{c}_{z}$ as a function of temperature in \ce{BaTiO3}. At each temperature, the coercive field is obtained by averaging over 20 trajectories initialized at different velocities. The corresponding standard deviation is given as shaded region around the average value. The absolute values of both the positive and negative coercive fields are shown. \textbf{d}, Polarization density $P_z/\Omega$ and electric field $E_z$ as a function of time in \ce{BaTiO3} at 300~K. $\tau$ denotes the period of the electric field.}
    \label{fig:results_BaTiO3}
\end{figure}

We apply our method to study dynamic ferroelectric properties of \ce{BaTiO3} perovskite. We consider the tetragonal phase, which is stable at room temperature, and train a ML model for \ce{BaTiO3} using 75 frames of 135 atoms each extracted from active learning dynamics. We start by calculating the ferroelectric hysteresis at zero temperature and illustrate the results in Fig.\ \ref{fig:results_BaTiO3}a. Each point of the hysteresis is obtained by performing a structural relaxation under an electric field along $\mathbf z$, which is varied following a sinusoidal behavior [cf.\ equation \eqref{eq:efield_t}]. To validate our result, we first examine the hysteresis for the 135-atom supercell through DFT calculations with finite electric fields. In this case, the values of polarization obtained do not all belong to the same Berry phase branch. Hence, for comparison with our ML results, we fold the DFT values of polarization to the Berry phase branch obtained with our ML model. As illustrated in Fig.\ \ref{fig:results_BaTiO3}a, we find a remarkable agreement between the ML and DFT ferroelectric hysteresis, which validates our ML model for \ce{BaTiO3}. For further validation, we perform the same analysis for a 40-atom supercell. 
As illustrated in Fig.\ \ref{fig:results_BaTiO3}a, our ML model trained solely on the 135-atoms supercell is able to reproduce the DFT hysteresis obtained for the 40-atom supercell, despite differences in size. 
This highlights the capabilities of the ML model in generalizing and capturing polarization behavior across simulations of different size, and confirms the accuracy of our model.

Next, we assess the temperature effects on the ferroelectric response in \ce{BaTiO3} by performing MLMD in the NVT ensemble under the influence of sinusoidally varying electric field (cf.\ Fig.\ \ref{fig:results_BaTiO3}b). We use a 1080-atom supercell and show the hysteresis at 300~K in Fig.\ \ref{fig:results_BaTiO3}b. The coercive field at 300~K is noticeably smaller than that at zero temperature (cf.\ Fig.\ \ref{fig:results_BaTiO3}b), which is consistent with the activated nucleation mechanism of hysteresis. We further assess the temperature dependence of the coercive field for various temperatures, observing a decrease as temperature increases, as illustrated in Fig.\ \ref{fig:results_BaTiO3}c. The hysteresis loops at both zero and finite temperature exhibit symmetry with respect to the direction of the electric field. This reflects the correct description of the polarization being a conservative vector field, which is a key consequence of the differential learning approach. 

\begin{figure*}[t!]
    \centerline{\includegraphics[width=1.0\linewidth]{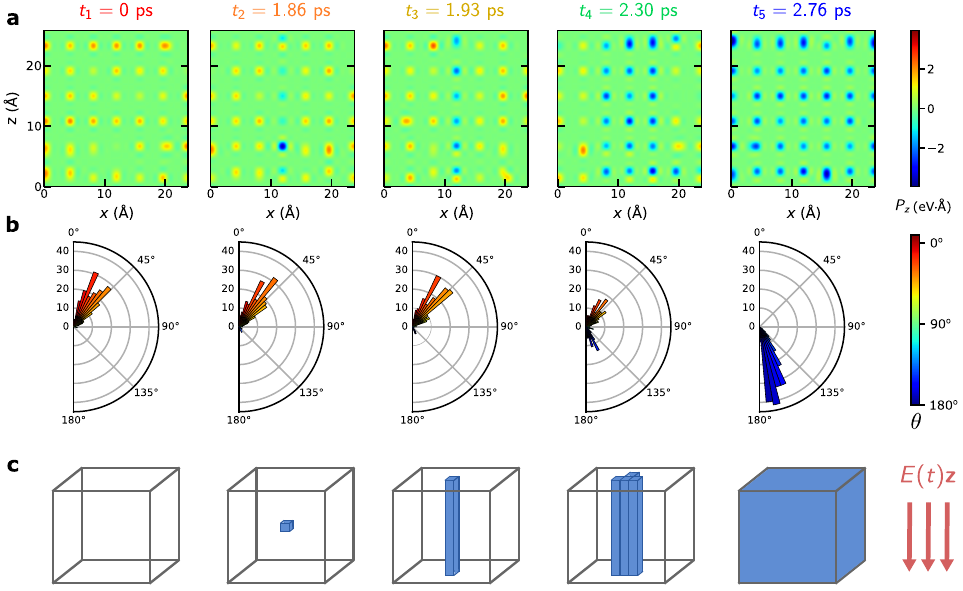}}
    \caption{\textbf{Dipole dynamics during the hysteresis transition in tetragonal \ce{BaTiO3} at room temperature}. \textbf{a}, Dipoles $P_z$ summed over planes in tetragonal \ce{BaTiO3} during the hysteresis transition from up to down polarization in the presence of a sinusoidal electric field in the $\mathbf z$ direction at 300~K. Dipoles are assigned for each 5-atom unit cell [cf.\ equation \eqref{eq:dipoles_unitcell}] and are located at the coordinates of Ti atoms. Close to the transition point ($t_1$), the dipoles are oriented in the up direction and are fluctuating due to temperature. While gradually decreasing the electric field, a single unit cell switches its polarization from up to down  ($t_2$). This is followed by an expansion of this down-polarized region along $\mathbf z$ ($t_3$), forming a one-dimensional domain line. Next, neighboring one-dimensional domain lines flip their polarization either along $\mathbf x$ or $\mathbf y$ ($t_4$), until the entire system is down polarized ($t_5$). A supercell of 1080 atoms is used. The time signatures $t_i$ are also shown in Fig.\ \ref{fig:results_BaTiO3}b. \textbf{b}, Histograms of the polar angle $\theta$ of the dipoles relative to the $\mathbf z$ axis during the hysteresis transition from up to down polarization. \textbf{c} Mechanism of polarization switching: \textit{(i)} nucleation of a polarization-flipped unit cell, \textit{(ii)} expansion to a one-dimensional domain line, \textit{(iii)} formation of neighboring one-dimensional domain lines along either the $\mathbf x$ or $\mathbf y$ direction, and \textit{(iv)} domain expansion through the entire system.}
    \label{fig:domain_hysteresis}
\end{figure*}

It is of interest to investigate the dynamics of ferroelectric dipoles in MD under an electric field, as these can reveal physical insight on ferroelectric domain formation and motion. In particular, we study the ferroelectric switching during the hysteresis in \ce{BaTiO3}. As illustrated in Fig.\ \ref{fig:domain_hysteresis}a, starting from an initial configuration with up polarization, the gradual decrease of the electric field along $\mathbf z$ induces the nucleation of a down polarized unit cell. The down-polarized region then propagates along the $\mathbf z$ direction, thereby creating one-dimensional domain line. The expansion along $\mathbf z$ is favored over the other Cartesian directions as diagonal Born charges $Z_{zz}^*$ are greater than the off-diagonal ones. 
Next, neighboring one-dimensional domain lines flip their polarization, until the entire system is down-polarized. In this process, expansion along $\mathbf x$ and $\mathbf y$ are equally probable, due to their equivalence in tetragonal \ce{BaTiO3}. 
The larger the down-polarized region is, the faster its expansion becomes, as its surface in the $xy$ plane encompasses more neighboring domain lines. This entire process for a supercell of 14.6 nm$^{3}$ happens in about 3 ps, which we visualize through time signatures that are shown in Fig.\ \ref{fig:results_BaTiO3}b. In Fig.\ \ref{fig:domain_hysteresis}b, we illustrate the corresponding dynamics of the polar angles formed by the dipole with the $\mathbf z$ axis throughout the polarization switching. In Fig.\ \ref{fig:domain_hysteresis}c, we provide a sketch of the underlying microscopic mechanism of nucleation and grown governing the ferroelectric domain switching.

\vspace{0.3cm} \noindent {\large \textbf{Discussion}}\\
In summary, we introduced a framework for learning the generalized potential energy and related response quantities to external fields within a unified ML model. This has several advantages. First, the response properties, obtained by differentiating the generalized potential energy with respect to the inputs of the model, by construction obey physical symmetries and conservation laws. Our approach ensures momentum conservation, the acoustic sum rule for Born charges, the polarization being a conservative vector field, and the electric enthalpy conservation in machine-learning molecular dynamics and in cyclic adiabatic evolutions involving changes of the electric field. Second, the differential Sobolev training approach allows for a richer set of training targets to be used in learning the generalized potential energy as a function of atomic coordinates and arbitrary parameters. We deploy the methods to enable simulating molecular dynamics of extended systems under the influence of applied electric fields. To this aim, we develop and implement a unified equivariant neural network model that learns the electric enthalpy and predict therefrom polarization, Born charges, and polarizability in addition to forces and stress. The model is based on an equivariant local description of the atomic environments, which offers advantages in accuracy, data efficiency, and scalability. We applied our model to determine the vibrational and dielectric properties of $\alpha$-\ce{SiO2}, finding excellent agreement with reference DFT and DFPT results. This demonstrates that our formulation can be used to perform large-scale simulations under finite electric fields. Next, we used our method to calculate temperature-dependent ferroelectric properties of \ce{BaTiO3}, and to reveal the microscopic mechanisms of domain wall nucleation and motion using large-scale molecular dynamics under applied electric fields at first-principles accuracy. We found that polarization switching starts from a nucleation of dipole reversal in a single 5-atom unit cell, followed by expansion to a one-dimensional domain line. Next, neighboring one-dimensional domain lines form along the $\mathbf x$ and $\mathbf y$ directions, until the entire supercell switches its polarization in just a few picoseconds.

The notable advance of our formulation is the ability to predict with first-principles accuracy the response properties of extended systems of much larger size than is possible with electronic structure methods, while also ensuring excellent convergence of sampling time correlations over long simulations duration. This opens possibilities to study dielectric, spectral, and ferroelectric properties of previously intractable complex systems with defects and disorder. We remark that our model is based on a local representation of the atomic environments and, therefore, long-range dipole-dipole interactions are not guaranteed to be fully captured. While this remains to be investigated quantitatively, we note that such long-range interactions are typically mitigated by screening effects in extended homogeneous systems such as those considered in this work. Furthermore, in the interest of comparing the first-principles results with experimental data, it may be important to access very large size and time scales of simulations. For instance, the values of electric field that we applied for observing the ferroelectric hysteresis in \ce{BaTiO3} exceed those applied experimentally or in previous studies \cite{shin2007nature,liu2016nature}. In relation to experiments, given that hysteresis is driven by irreversible domain nucleation and growth processes, our ideal crystal geometry may not represent multiple ferroelectric domains or defects present in experimental samples, which would allow for lower coercive electric field values for polarization switching. Additionally, strain variations and surface effects in thin-film samples may significantly influence the magnitude of the coercive field. Finally, much lower frequencies of the applied electric field are typically used in experiments or in previous studies \cite{shin2007nature,liu2016nature}, which reduce the magnitude of the coercive field, as the system would have more time to nucleate a local polarization reversal. The investigation of such effects, which would certainly be intractable with standard quantum mechanical methods, is feasible within our method and is left for future studies. 

In the context of ferroelectric dynamics, our approach enables first-principles real-time simulations of ferroelectric switching, where polarization and Born charges are treated within the modern theory of polarization, going beyond previous models based on empirical parameters \cite{shin2007nature,liu2016nature} or uncontrolled approximations for Born charges \cite{akbarian2019PCCP,deguchi2023PSS}. This formulation enables accurate modeling of complex structures, such as vacancies, polarons, and polarization vortices in heterogeneous geometries, for which simple models are not applicable. In addition, our method can be applied for modeling an entire nanoscale ferroelectric device or nanoparticle, which would offer tremendous advantages for technological advancements.
Overall, our work offers a promising new direction in using machine learning techniques to accelerate the investigation of dielectric, vibrational and ferroelectric properties of complex materials, including crystalline, disordered, and liquid systems. Finally, ideas introduced in this work generalize readily to differentiable strategies for efficient learning of a wide variety of generalized potentials, such as the free energy and grand canonical potential, and to higher order responses, such as piezoelectric and magnetostrictive coefficients. Broadly,  paves the way to materials design and understanding through machine learning methods for modeling responses under external fields while satisfying exact physical symmetries and conservation laws.

\bibliography{bibliography}

\providecommand{\noopsort}[1]{}\providecommand{\singleletter}[1]{#1}%
\begin{thebibliography}{50}%
\makeatletter
\providecommand \@ifxundefined [1]{%
 \@ifx{#1\undefined}
}%
\providecommand \@ifnum [1]{%
 \ifnum #1\expandafter \@firstoftwo
 \else \expandafter \@secondoftwo
 \fi
}%
\providecommand \@ifx [1]{%
 \ifx #1\expandafter \@firstoftwo
 \else \expandafter \@secondoftwo
 \fi
}%
\providecommand \natexlab [1]{#1}%
\providecommand \enquote  [1]{``#1''}%
\providecommand \bibnamefont  [1]{#1}%
\providecommand \bibfnamefont [1]{#1}%
\providecommand \citenamefont [1]{#1}%
\providecommand \href@noop [0]{\@secondoftwo}%
\providecommand \href [0]{\begingroup \@sanitize@url \@href}%
\providecommand \@href[1]{\@@startlink{#1}\@@href}%
\providecommand \@@href[1]{\endgroup#1\@@endlink}%
\providecommand \@sanitize@url [0]{\catcode `\\12\catcode `\$12\catcode
  `\&12\catcode `\#12\catcode `\^12\catcode `\_12\catcode `\%12\relax}%
\providecommand \@@startlink[1]{}%
\providecommand \@@endlink[0]{}%
\providecommand \url  [0]{\begingroup\@sanitize@url \@url }%
\providecommand \@url [1]{\endgroup\@href {#1}{\urlprefix }}%
\providecommand \urlprefix  [0]{URL }%
\providecommand \Eprint [0]{\href }%
\providecommand \doibase [0]{https://doi.org/}%
\providecommand \selectlanguage [0]{\@gobble}%
\providecommand \bibinfo  [0]{\@secondoftwo}%
\providecommand \bibfield  [0]{\@secondoftwo}%
\providecommand \translation [1]{[#1]}%
\providecommand \BibitemOpen [0]{}%
\providecommand \bibitemStop [0]{}%
\providecommand \bibitemNoStop [0]{.\EOS\space}%
\providecommand \EOS [0]{\spacefactor3000\relax}%
\providecommand \BibitemShut  [1]{\csname bibitem#1\endcsname}%
\let\auto@bib@innerbib\@empty
\bibitem [{\citenamefont {Baroni}\ \emph {et~al.}(2001)\citenamefont {Baroni},
  \citenamefont {de~Gironcoli}, \citenamefont {Dal~Corso},\ and\ \citenamefont
  {Giannozzi}}]{baroni2001RMP}%
  \BibitemOpen
  \bibfield  {author} {\bibinfo {author} {\bibfnamefont {S.}~\bibnamefont
  {Baroni}}, \bibinfo {author} {\bibfnamefont {S.}~\bibnamefont
  {de~Gironcoli}}, \bibinfo {author} {\bibfnamefont {A.}~\bibnamefont
  {Dal~Corso}},\ and\ \bibinfo {author} {\bibfnamefont {P.}~\bibnamefont
  {Giannozzi}},\ }\bibfield  {title} {\bibinfo {title} {Phonons and related
  crystal properties from density-functional perturbation theory},\ }\href
  {https://doi.org/10.1103/RevModPhys.73.515} {\bibfield  {journal} {\bibinfo
  {journal} {Rev. Mod. Phys.}\ }\textbf {\bibinfo {volume} {73}},\ \bibinfo
  {pages} {515} (\bibinfo {year} {2001})}\BibitemShut {NoStop}%
\bibitem [{\citenamefont {Gonze}\ and\ \citenamefont
  {Lee}(1997)}]{gonze1997PRB_2}%
  \BibitemOpen
  \bibfield  {author} {\bibinfo {author} {\bibfnamefont {X.}~\bibnamefont
  {Gonze}}\ and\ \bibinfo {author} {\bibfnamefont {C.}~\bibnamefont {Lee}},\
  }\bibfield  {title} {\bibinfo {title} {Dynamical matrices, {Born} effective
  charges, dielectric permittivity tensors, and interatomic force constants
  from density-functional perturbation theory},\ }\href
  {https://doi.org/10.1103/PhysRevB.55.10355} {\bibfield  {journal} {\bibinfo
  {journal} {Phys. Rev. B}\ }\textbf {\bibinfo {volume} {55}},\ \bibinfo
  {pages} {10355} (\bibinfo {year} {1997})}\BibitemShut {NoStop}%
\bibitem [{\citenamefont {Batzner}\ \emph {et~al.}(2022)\citenamefont
  {Batzner}, \citenamefont {Musaelian}, \citenamefont {Sun}, \citenamefont
  {Geiger}, \citenamefont {Mailoa}, \citenamefont {Kornbluth}, \citenamefont
  {Molinari}, \citenamefont {Smidt},\ and\ \citenamefont
  {Kozinsky}}]{batzner2023NC}%
  \BibitemOpen
  \bibfield  {author} {\bibinfo {author} {\bibfnamefont {S.}~\bibnamefont
  {Batzner}}, \bibinfo {author} {\bibfnamefont {A.}~\bibnamefont {Musaelian}},
  \bibinfo {author} {\bibfnamefont {L.}~\bibnamefont {Sun}}, \bibinfo {author}
  {\bibfnamefont {M.}~\bibnamefont {Geiger}}, \bibinfo {author} {\bibfnamefont
  {J.~P.}\ \bibnamefont {Mailoa}}, \bibinfo {author} {\bibfnamefont
  {M.}~\bibnamefont {Kornbluth}}, \bibinfo {author} {\bibfnamefont
  {N.}~\bibnamefont {Molinari}}, \bibinfo {author} {\bibfnamefont {T.~E.}\
  \bibnamefont {Smidt}},\ and\ \bibinfo {author} {\bibfnamefont
  {B.}~\bibnamefont {Kozinsky}},\ }\bibfield  {title} {\bibinfo {title}
  {{E}(3)-equivariant graph neural networks for data-efficient and accurate
  interatomic potentials},\ }\href {https://doi.org/10.1038/s41467-022-29939-5}
  {\bibfield  {journal} {\bibinfo  {journal} {Nat. Commun.}\ }\textbf {\bibinfo
  {volume} {13}},\ \bibinfo {pages} {2453} (\bibinfo {year}
  {2022})}\BibitemShut {NoStop}%
\bibitem [{\citenamefont {Musaelian}\ \emph {et~al.}(2023)\citenamefont
  {Musaelian}, \citenamefont {Batzner}, \citenamefont {Johansson},
  \citenamefont {Sun}, \citenamefont {Owen}, \citenamefont {Kornbluth},\ and\
  \citenamefont {Kozinsky}}]{musaelian2023NC}%
  \BibitemOpen
  \bibfield  {author} {\bibinfo {author} {\bibfnamefont {A.}~\bibnamefont
  {Musaelian}}, \bibinfo {author} {\bibfnamefont {S.}~\bibnamefont {Batzner}},
  \bibinfo {author} {\bibfnamefont {A.}~\bibnamefont {Johansson}}, \bibinfo
  {author} {\bibfnamefont {L.}~\bibnamefont {Sun}}, \bibinfo {author}
  {\bibfnamefont {C.~J.}\ \bibnamefont {Owen}}, \bibinfo {author}
  {\bibfnamefont {M.}~\bibnamefont {Kornbluth}},\ and\ \bibinfo {author}
  {\bibfnamefont {B.}~\bibnamefont {Kozinsky}},\ }\bibfield  {title} {\bibinfo
  {title} {Learning local equivariant representations for large-scale atomistic
  dynamics},\ }\href {https://doi.org/10.1038/s41467-023-36329-y} {\bibfield
  {journal} {\bibinfo  {journal} {Nat. Commun.}\ }\textbf {\bibinfo {volume}
  {14}},\ \bibinfo {pages} {579} (\bibinfo {year} {2023})}\BibitemShut
  {NoStop}%
\bibitem [{\citenamefont {Batatia}\ \emph {et~al.}(2022)\citenamefont
  {Batatia}, \citenamefont {Batzner}, \citenamefont {Kov{\'a}cs}, \citenamefont
  {Musaelian}, \citenamefont {Simm}, \citenamefont {Drautz}, \citenamefont
  {Ortner}, \citenamefont {Kozinsky},\ and\ \citenamefont
  {Cs{\'a}nyi}}]{batatia2022arxiv}%
  \BibitemOpen
  \bibfield  {author} {\bibinfo {author} {\bibfnamefont {I.}~\bibnamefont
  {Batatia}}, \bibinfo {author} {\bibfnamefont {S.}~\bibnamefont {Batzner}},
  \bibinfo {author} {\bibfnamefont {D.~P.}\ \bibnamefont {Kov{\'a}cs}},
  \bibinfo {author} {\bibfnamefont {A.}~\bibnamefont {Musaelian}}, \bibinfo
  {author} {\bibfnamefont {G.~N.}\ \bibnamefont {Simm}}, \bibinfo {author}
  {\bibfnamefont {R.}~\bibnamefont {Drautz}}, \bibinfo {author} {\bibfnamefont
  {C.}~\bibnamefont {Ortner}}, \bibinfo {author} {\bibfnamefont
  {B.}~\bibnamefont {Kozinsky}},\ and\ \bibinfo {author} {\bibfnamefont
  {G.}~\bibnamefont {Cs{\'a}nyi}},\ }\bibfield  {title} {\bibinfo {title} {The
  design space of e (3)-equivariant atom-centered interatomic potentials},\
  }\href {https://arxiv.org/abs/2205.06643} {\bibfield  {journal} {\bibinfo
  {journal} {arXiv:2205.06643}\ } (\bibinfo {year} {2022})}\BibitemShut
  {NoStop}%
\bibitem [{\citenamefont {Nigam}\ \emph {et~al.}(2022)\citenamefont {Nigam},
  \citenamefont {Willatt},\ and\ \citenamefont {Ceriotti}}]{nigam2022JCP}%
  \BibitemOpen
  \bibfield  {author} {\bibinfo {author} {\bibfnamefont {J.}~\bibnamefont
  {Nigam}}, \bibinfo {author} {\bibfnamefont {M.~J.}\ \bibnamefont {Willatt}},\
  and\ \bibinfo {author} {\bibfnamefont {M.}~\bibnamefont {Ceriotti}},\
  }\bibfield  {title} {\bibinfo {title} {{Equivariant representations for
  molecular Hamiltonians and N-center atomic-scale properties}},\ }\href
  {https://doi.org/10.1063/5.0072784} {\bibfield  {journal} {\bibinfo
  {journal} {J. Chem. Phys.}\ }\textbf {\bibinfo {volume} {156}},\ \bibinfo
  {pages} {014115} (\bibinfo {year} {2022})}\BibitemShut {NoStop}%
\bibitem [{\citenamefont {Geiger}\ and\ \citenamefont
  {Smidt}(2022)}]{geiger2022e3nn}%
  \BibitemOpen
  \bibfield  {author} {\bibinfo {author} {\bibfnamefont {M.}~\bibnamefont
  {Geiger}}\ and\ \bibinfo {author} {\bibfnamefont {T.}~\bibnamefont {Smidt}},\
  }\bibfield  {title} {\bibinfo {title} {e3nn: Euclidean neural networks},\
  }\href {https://arxiv.org/abs/2207.09453} {\bibfield  {journal} {\bibinfo
  {journal} {arXiv:2207.09453}\ } (\bibinfo {year} {2022})}\BibitemShut
  {NoStop}%
\bibitem [{\citenamefont {Neumann}(1983)}]{neumann1983MP}%
  \BibitemOpen
  \bibfield  {author} {\bibinfo {author} {\bibfnamefont {M.}~\bibnamefont
  {Neumann}},\ }\bibfield  {title} {\bibinfo {title} {Dipole moment fluctuation
  formulas in computer simulations of polar systems},\ }\href@noop {}
  {\bibfield  {journal} {\bibinfo  {journal} {Molecular Physics}\ }\textbf
  {\bibinfo {volume} {50}},\ \bibinfo {pages} {841} (\bibinfo {year}
  {1983})}\BibitemShut {NoStop}%
\bibitem [{\citenamefont {Neumann}\ and\ \citenamefont
  {Steinhauser}(1983)}]{neumann1983CPL}%
  \BibitemOpen
  \bibfield  {author} {\bibinfo {author} {\bibfnamefont {M.}~\bibnamefont
  {Neumann}}\ and\ \bibinfo {author} {\bibfnamefont {O.}~\bibnamefont
  {Steinhauser}},\ }\bibfield  {title} {\bibinfo {title} {On the calculation of
  the frequency-dependent dielectric constant in computer simulations},\ }\href
  {https://doi.org/https://doi.org/10.1016/0009-2614(83)87455-7} {\bibfield
  {journal} {\bibinfo  {journal} {Chem. Phys. Lett.}\ }\textbf {\bibinfo
  {volume} {102}},\ \bibinfo {pages} {508} (\bibinfo {year}
  {1983})}\BibitemShut {NoStop}%
\bibitem [{\citenamefont {Shin}\ \emph {et~al.}(2007)\citenamefont {Shin},
  \citenamefont {Grinberg}, \citenamefont {Chen},\ and\ \citenamefont
  {Rappe}}]{shin2007nature}%
  \BibitemOpen
  \bibfield  {author} {\bibinfo {author} {\bibfnamefont {Y.-H.}\ \bibnamefont
  {Shin}}, \bibinfo {author} {\bibfnamefont {I.}~\bibnamefont {Grinberg}},
  \bibinfo {author} {\bibfnamefont {I.-W.}\ \bibnamefont {Chen}},\ and\
  \bibinfo {author} {\bibfnamefont {A.~M.}\ \bibnamefont {Rappe}},\ }\bibfield
  {title} {\bibinfo {title} {Nucleation and growth mechanism of ferroelectric
  domain-wall motion},\ }\href {https://doi.org/10.1038/nature06165} {\bibfield
   {journal} {\bibinfo  {journal} {Nature}\ }\textbf {\bibinfo {volume}
  {449}},\ \bibinfo {pages} {881} (\bibinfo {year} {2007})}\BibitemShut
  {NoStop}%
\bibitem [{\citenamefont {Liu}\ \emph {et~al.}(2016)\citenamefont {Liu},
  \citenamefont {Grinberg},\ and\ \citenamefont {Rappe}}]{liu2016nature}%
  \BibitemOpen
  \bibfield  {author} {\bibinfo {author} {\bibfnamefont {S.}~\bibnamefont
  {Liu}}, \bibinfo {author} {\bibfnamefont {I.}~\bibnamefont {Grinberg}},\ and\
  \bibinfo {author} {\bibfnamefont {A.~M.}\ \bibnamefont {Rappe}},\ }\bibfield
  {title} {\bibinfo {title} {Intrinsic ferroelectric switching from first
  principles},\ }\href {https://doi.org/10.1038/nature18286} {\bibfield
  {journal} {\bibinfo  {journal} {Nature}\ }\textbf {\bibinfo {volume} {534}},\
  \bibinfo {pages} {360} (\bibinfo {year} {2016})}\BibitemShut {NoStop}%
\bibitem [{\citenamefont {Resta}(1994)}]{resta1994RMP}%
  \BibitemOpen
  \bibfield  {author} {\bibinfo {author} {\bibfnamefont {R.}~\bibnamefont
  {Resta}},\ }\bibfield  {title} {\bibinfo {title} {Macroscopic polarization in
  crystalline dielectrics: the geometric phase approach},\ }\href
  {https://doi.org/10.1103/RevModPhys.66.899} {\bibfield  {journal} {\bibinfo
  {journal} {Rev. Mod. Phys.}\ }\textbf {\bibinfo {volume} {66}},\ \bibinfo
  {pages} {899} (\bibinfo {year} {1994})}\BibitemShut {NoStop}%
\bibitem [{\citenamefont {King-Smith}\ and\ \citenamefont
  {Vanderbilt}(1993)}]{kingsmith1993PRB}%
  \BibitemOpen
  \bibfield  {author} {\bibinfo {author} {\bibfnamefont {R.~D.}\ \bibnamefont
  {King-Smith}}\ and\ \bibinfo {author} {\bibfnamefont {D.}~\bibnamefont
  {Vanderbilt}},\ }\bibfield  {title} {\bibinfo {title} {Theory of polarization
  of crystalline solids},\ }\href {https://doi.org/10.1103/PhysRevB.47.1651}
  {\bibfield  {journal} {\bibinfo  {journal} {Phys. Rev. B}\ }\textbf {\bibinfo
  {volume} {47}},\ \bibinfo {pages} {1651} (\bibinfo {year}
  {1993})}\BibitemShut {NoStop}%
\bibitem [{\citenamefont {Spaldin}(2012)}]{spaldin2012JSSC}%
  \BibitemOpen
  \bibfield  {author} {\bibinfo {author} {\bibfnamefont {N.~A.}\ \bibnamefont
  {Spaldin}},\ }\bibfield  {title} {\bibinfo {title} {A beginner's guide to the
  modern theory of polarization},\ }\href
  {https://doi.org/https://doi.org/10.1016/j.jssc.2012.05.010} {\bibfield
  {journal} {\bibinfo  {journal} {J. Solid State Chem.}\ }\textbf {\bibinfo
  {volume} {195}},\ \bibinfo {pages} {2} (\bibinfo {year} {2012})}\BibitemShut
  {NoStop}%
\bibitem [{\citenamefont {Nunes}\ and\ \citenamefont
  {Gonze}(2001)}]{nunes2001PRB}%
  \BibitemOpen
  \bibfield  {author} {\bibinfo {author} {\bibfnamefont {R.~W.}\ \bibnamefont
  {Nunes}}\ and\ \bibinfo {author} {\bibfnamefont {X.}~\bibnamefont {Gonze}},\
  }\bibfield  {title} {\bibinfo {title} {Berry-phase treatment of the
  homogeneous electric field perturbation in insulators},\ }\href
  {https://doi.org/10.1103/PhysRevB.63.155107} {\bibfield  {journal} {\bibinfo
  {journal} {Phys. Rev. B}\ }\textbf {\bibinfo {volume} {63}},\ \bibinfo
  {pages} {155107} (\bibinfo {year} {2001})}\BibitemShut {NoStop}%
\bibitem [{\citenamefont {Umari}\ and\ \citenamefont
  {Pasquarello}(2002)}]{umari2002PRL}%
  \BibitemOpen
  \bibfield  {author} {\bibinfo {author} {\bibfnamefont {P.}~\bibnamefont
  {Umari}}\ and\ \bibinfo {author} {\bibfnamefont {A.}~\bibnamefont
  {Pasquarello}},\ }\bibfield  {title} {\bibinfo {title} {Ab initio molecular
  dynamics in a finite homogeneous electric field},\ }\href
  {https://doi.org/10.1103/PhysRevLett.89.157602} {\bibfield  {journal}
  {\bibinfo  {journal} {Phys. Rev. Lett.}\ }\textbf {\bibinfo {volume} {89}},\
  \bibinfo {pages} {157602} (\bibinfo {year} {2002})}\BibitemShut {NoStop}%
\bibitem [{\citenamefont {Christensen}\ \emph {et~al.}(2019)\citenamefont
  {Christensen}, \citenamefont {Faber},\ and\ \citenamefont {von
  Lilienfeld}}]{christensen2019JCP}%
  \BibitemOpen
  \bibfield  {author} {\bibinfo {author} {\bibfnamefont {A.~S.}\ \bibnamefont
  {Christensen}}, \bibinfo {author} {\bibfnamefont {F.~A.}\ \bibnamefont
  {Faber}},\ and\ \bibinfo {author} {\bibfnamefont {O.~A.}\ \bibnamefont {von
  Lilienfeld}},\ }\bibfield  {title} {\bibinfo {title} {{Operators in quantum
  machine learning: Response properties in chemical space}},\ }\href
  {https://doi.org/10.1063/1.5053562} {\bibfield  {journal} {\bibinfo
  {journal} {J. Chem. Phys.}\ }\textbf {\bibinfo {volume} {150}},\ \bibinfo
  {pages} {064105} (\bibinfo {year} {2019})}\BibitemShut {NoStop}%
\bibitem [{\citenamefont {Veit}\ \emph {et~al.}(2020)\citenamefont {Veit},
  \citenamefont {Wilkins}, \citenamefont {Yang}, \citenamefont {DiStasio},\
  and\ \citenamefont {Ceriotti}}]{veit2020JCP}%
  \BibitemOpen
  \bibfield  {author} {\bibinfo {author} {\bibfnamefont {M.}~\bibnamefont
  {Veit}}, \bibinfo {author} {\bibfnamefont {D.~M.}\ \bibnamefont {Wilkins}},
  \bibinfo {author} {\bibfnamefont {Y.}~\bibnamefont {Yang}}, \bibinfo {author}
  {\bibfnamefont {J.}~\bibnamefont {DiStasio}, \bibfnamefont {Robert~A.}},\
  and\ \bibinfo {author} {\bibfnamefont {M.}~\bibnamefont {Ceriotti}},\
  }\bibfield  {title} {\bibinfo {title} {{Predicting molecular dipole moments
  by combining atomic partial charges and atomic dipoles}},\ }\href
  {https://doi.org/10.1063/5.0009106} {\bibfield  {journal} {\bibinfo
  {journal} {J. Chem. Phys.}\ }\textbf {\bibinfo {volume} {153}},\ \bibinfo
  {pages} {024113} (\bibinfo {year} {2020})}\BibitemShut {NoStop}%
\bibitem [{\citenamefont {Krishnamoorthy}\ \emph {et~al.}(2021)\citenamefont
  {Krishnamoorthy}, \citenamefont {Nomura}, \citenamefont {Baradwaj},
  \citenamefont {Shimamura}, \citenamefont {Rajak}, \citenamefont {Mishra},
  \citenamefont {Fukushima}, \citenamefont {Shimojo}, \citenamefont {Kalia},
  \citenamefont {Nakano},\ and\ \citenamefont
  {Vashishta}}]{krishnamoorthy2021PRL}%
  \BibitemOpen
  \bibfield  {author} {\bibinfo {author} {\bibfnamefont {A.}~\bibnamefont
  {Krishnamoorthy}}, \bibinfo {author} {\bibfnamefont {K.-i.}\ \bibnamefont
  {Nomura}}, \bibinfo {author} {\bibfnamefont {N.}~\bibnamefont {Baradwaj}},
  \bibinfo {author} {\bibfnamefont {K.}~\bibnamefont {Shimamura}}, \bibinfo
  {author} {\bibfnamefont {P.}~\bibnamefont {Rajak}}, \bibinfo {author}
  {\bibfnamefont {A.}~\bibnamefont {Mishra}}, \bibinfo {author} {\bibfnamefont
  {S.}~\bibnamefont {Fukushima}}, \bibinfo {author} {\bibfnamefont
  {F.}~\bibnamefont {Shimojo}}, \bibinfo {author} {\bibfnamefont
  {R.}~\bibnamefont {Kalia}}, \bibinfo {author} {\bibfnamefont
  {A.}~\bibnamefont {Nakano}},\ and\ \bibinfo {author} {\bibfnamefont
  {P.}~\bibnamefont {Vashishta}},\ }\bibfield  {title} {\bibinfo {title}
  {Dielectric constant of liquid water determined with neural network quantum
  molecular dynamics},\ }\href {https://doi.org/10.1103/PhysRevLett.126.216403}
  {\bibfield  {journal} {\bibinfo  {journal} {Phys. Rev. Lett.}\ }\textbf
  {\bibinfo {volume} {126}},\ \bibinfo {pages} {216403} (\bibinfo {year}
  {2021})}\BibitemShut {NoStop}%
\bibitem [{\citenamefont {Gastegger}\ \emph {et~al.}(2021)\citenamefont
  {Gastegger}, \citenamefont {Schütt},\ and\ \citenamefont
  {Müller}}]{gastegger2021CS}%
  \BibitemOpen
  \bibfield  {author} {\bibinfo {author} {\bibfnamefont {M.}~\bibnamefont
  {Gastegger}}, \bibinfo {author} {\bibfnamefont {K.~T.}\ \bibnamefont
  {Schütt}},\ and\ \bibinfo {author} {\bibfnamefont {K.-R.}\ \bibnamefont
  {Müller}},\ }\bibfield  {title} {\bibinfo {title} {Machine learning of
  solvent effects on molecular spectra and reactions},\ }\href
  {https://doi.org/10.1039/D1SC02742E} {\bibfield  {journal} {\bibinfo
  {journal} {Chem. Sci.}\ }\textbf {\bibinfo {volume} {12}},\ \bibinfo {pages}
  {11473} (\bibinfo {year} {2021})}\BibitemShut {NoStop}%
\bibitem [{\citenamefont {Staacke}\ \emph {et~al.}(2022)\citenamefont
  {Staacke}, \citenamefont {Wengert}, \citenamefont {Kunkel}, \citenamefont
  {Csányi}, \citenamefont {Reuter},\ and\ \citenamefont
  {Margraf}}]{staackeMLST2022}%
  \BibitemOpen
  \bibfield  {author} {\bibinfo {author} {\bibfnamefont {C.~G.}\ \bibnamefont
  {Staacke}}, \bibinfo {author} {\bibfnamefont {S.}~\bibnamefont {Wengert}},
  \bibinfo {author} {\bibfnamefont {C.}~\bibnamefont {Kunkel}}, \bibinfo
  {author} {\bibfnamefont {G.}~\bibnamefont {Csányi}}, \bibinfo {author}
  {\bibfnamefont {K.}~\bibnamefont {Reuter}},\ and\ \bibinfo {author}
  {\bibfnamefont {J.~T.}\ \bibnamefont {Margraf}},\ }\bibfield  {title}
  {\bibinfo {title} {Kernel charge equilibration: efficient and accurate
  prediction of molecular dipole moments with a machine-learning enhanced
  electron density model},\ }\href {https://doi.org/10.1088/2632-2153/ac568d}
  {\bibfield  {journal} {\bibinfo  {journal} {Mach. Learn.: Sci. Technol.}\
  }\textbf {\bibinfo {volume} {3}},\ \bibinfo {pages} {015032} (\bibinfo {year}
  {2022})}\BibitemShut {NoStop}%
\bibitem [{\citenamefont {Gigli}\ \emph {et~al.}(2022)\citenamefont {Gigli},
  \citenamefont {Veit}, \citenamefont {Kotiuga}, \citenamefont {Pizzi},
  \citenamefont {Marzari},\ and\ \citenamefont {Ceriotti}}]{gigli2022npj}%
  \BibitemOpen
  \bibfield  {author} {\bibinfo {author} {\bibfnamefont {L.}~\bibnamefont
  {Gigli}}, \bibinfo {author} {\bibfnamefont {M.}~\bibnamefont {Veit}},
  \bibinfo {author} {\bibfnamefont {M.}~\bibnamefont {Kotiuga}}, \bibinfo
  {author} {\bibfnamefont {G.}~\bibnamefont {Pizzi}}, \bibinfo {author}
  {\bibfnamefont {N.}~\bibnamefont {Marzari}},\ and\ \bibinfo {author}
  {\bibfnamefont {M.}~\bibnamefont {Ceriotti}},\ }\bibfield  {title} {\bibinfo
  {title} {Thermodynamics and dielectric response of {BaTiO$_3$} by data-driven
  modeling},\ }\href {https://doi.org/10.1038/s41524-022-00845-0} {\bibfield
  {journal} {\bibinfo  {journal} {npj Comput. Mater.}\ }\textbf {\bibinfo
  {volume} {8}},\ \bibinfo {pages} {209} (\bibinfo {year} {2022})}\BibitemShut
  {NoStop}%
\bibitem [{\citenamefont {Schütt}\ \emph {et~al.}(2021)\citenamefont
  {Schütt}, \citenamefont {Unke},\ and\ \citenamefont
  {Gastegger}}]{schutt2021ICML}%
  \BibitemOpen
  \bibfield  {author} {\bibinfo {author} {\bibfnamefont {K.}~\bibnamefont
  {Schütt}}, \bibinfo {author} {\bibfnamefont {O.}~\bibnamefont {Unke}},\ and\
  \bibinfo {author} {\bibfnamefont {M.}~\bibnamefont {Gastegger}},\ }\bibfield
  {title} {\bibinfo {title} {Equivariant message passing for the prediction of
  tensorial properties and molecular spectra},\ }\href
  {https://proceedings.mlr.press/v139/schutt21a.html} {\bibfield  {journal}
  {\bibinfo  {journal} {Int. Conf. Mach. Learn.}\ }\textbf {\bibinfo {volume}
  {139}},\ \bibinfo {pages} {9377} (\bibinfo {year} {2021})}\BibitemShut
  {NoStop}%
\bibitem [{\citenamefont {Schienbein}(2023)}]{schienbein2023JCTC}%
  \BibitemOpen
  \bibfield  {author} {\bibinfo {author} {\bibfnamefont {P.}~\bibnamefont
  {Schienbein}},\ }\bibfield  {title} {\bibinfo {title} {Spectroscopy from
  machine learning by accurately representing the atomic polar tensor},\ }\href
  {https://doi.org/10.1021/acs.jctc.2c00788} {\bibfield  {journal} {\bibinfo
  {journal} {J. Chem. Theory Comput.}\ }\textbf {\bibinfo {volume} {19}},\
  \bibinfo {pages} {705} (\bibinfo {year} {2023})}\BibitemShut {NoStop}%
\bibitem [{\citenamefont {Shao}\ \emph {et~al.}(2023)\citenamefont {Shao},
  \citenamefont {Paetow}, \citenamefont {Tuckerman},\ and\ \citenamefont
  {Pavanello}}]{shao2023NC}%
  \BibitemOpen
  \bibfield  {author} {\bibinfo {author} {\bibfnamefont {X.}~\bibnamefont
  {Shao}}, \bibinfo {author} {\bibfnamefont {L.}~\bibnamefont {Paetow}},
  \bibinfo {author} {\bibfnamefont {M.~E.}\ \bibnamefont {Tuckerman}},\ and\
  \bibinfo {author} {\bibfnamefont {M.}~\bibnamefont {Pavanello}},\ }\bibfield
  {title} {\bibinfo {title} {Machine learning electronic structure methods
  based on the one-electron reduced density matrix},\ }\href
  {https://doi.org/10.1038/s41467-023-41953-9} {\bibfield  {journal} {\bibinfo
  {journal} {Nat. Commun.}\ }\textbf {\bibinfo {volume} {14}} (\bibinfo {year}
  {2023})}\BibitemShut {NoStop}%
\bibitem [{\citenamefont {Zhang}\ \emph {et~al.}(2022)\citenamefont {Zhang},
  \citenamefont {Wang}, \citenamefont {Muniz}, \citenamefont {Panagiotopoulos},
  \citenamefont {Car},\ and\ \citenamefont {E}}]{zhang2022JCP}%
  \BibitemOpen
  \bibfield  {author} {\bibinfo {author} {\bibfnamefont {L.}~\bibnamefont
  {Zhang}}, \bibinfo {author} {\bibfnamefont {H.}~\bibnamefont {Wang}},
  \bibinfo {author} {\bibfnamefont {M.~C.}\ \bibnamefont {Muniz}}, \bibinfo
  {author} {\bibfnamefont {A.~Z.}\ \bibnamefont {Panagiotopoulos}}, \bibinfo
  {author} {\bibfnamefont {R.}~\bibnamefont {Car}},\ and\ \bibinfo {author}
  {\bibfnamefont {W.}~\bibnamefont {E}},\ }\bibfield  {title} {\bibinfo {title}
  {{A deep potential model with long-range electrostatic interactions}},\
  }\href {https://doi.org/10.1063/5.0083669} {\bibfield  {journal} {\bibinfo
  {journal} {J. Chem. Phys.}\ }\textbf {\bibinfo {volume} {156}},\ \bibinfo
  {pages} {124107} (\bibinfo {year} {2022})}\BibitemShut {NoStop}%
\bibitem [{\citenamefont {Joll}\ \emph {et~al.}(2024)\citenamefont {Joll},
  \citenamefont {Schienbein}, \citenamefont {Rosso},\ and\ \citenamefont
  {Blumberger}}]{joll2024arxiv}%
  \BibitemOpen
  \bibfield  {author} {\bibinfo {author} {\bibfnamefont {K.}~\bibnamefont
  {Joll}}, \bibinfo {author} {\bibfnamefont {P.}~\bibnamefont {Schienbein}},
  \bibinfo {author} {\bibfnamefont {K.~M.}\ \bibnamefont {Rosso}},\ and\
  \bibinfo {author} {\bibfnamefont {J.}~\bibnamefont {Blumberger}},\
  }\href@noop {} {\bibinfo {title} {Molecular dynamics simulation with finite
  electric fields using perturbed neural network potentials}} (\bibinfo {year}
  {2024}),\ \Eprint {https://arxiv.org/abs/2403.12319} {arXiv:2403.12319}
  \BibitemShut {NoStop}%
\bibitem [{\citenamefont {Shimizu}\ \emph {et~al.}(2023)\citenamefont
  {Shimizu}, \citenamefont {Otsuka}, \citenamefont {Hara}, \citenamefont
  {Minamitani},\ and\ \citenamefont {Watanabe}}]{shimizu2023STAM}%
  \BibitemOpen
  \bibfield  {author} {\bibinfo {author} {\bibfnamefont {K.}~\bibnamefont
  {Shimizu}}, \bibinfo {author} {\bibfnamefont {R.}~\bibnamefont {Otsuka}},
  \bibinfo {author} {\bibfnamefont {M.}~\bibnamefont {Hara}}, \bibinfo {author}
  {\bibfnamefont {E.}~\bibnamefont {Minamitani}},\ and\ \bibinfo {author}
  {\bibfnamefont {S.}~\bibnamefont {Watanabe}},\ }\bibfield  {title} {\bibinfo
  {title} {Prediction of {Born} effective charges using neural network to study
  ion migration under electric fields: applications to crystalline and
  amorphous {Li$_3$PO$_4$}},\ }\href
  {https://doi.org/10.1080/27660400.2023.2253135} {\bibfield  {journal}
  {\bibinfo  {journal} {Sci. Technol. Adv.}\ }\textbf {\bibinfo {volume} {3}},\
  \bibinfo {pages} {2253135} (\bibinfo {year} {2023})}\BibitemShut {NoStop}%
\bibitem [{\citenamefont {Choudhary}\ \emph {et~al.}(2020)\citenamefont
  {Choudhary}, \citenamefont {Garrity}, \citenamefont {Sharma}, \citenamefont
  {Biacchi}, \citenamefont {Hight~Walker},\ and\ \citenamefont
  {Tavazza}}]{choudhary2020npj}%
  \BibitemOpen
  \bibfield  {author} {\bibinfo {author} {\bibfnamefont {K.}~\bibnamefont
  {Choudhary}}, \bibinfo {author} {\bibfnamefont {K.~F.}\ \bibnamefont
  {Garrity}}, \bibinfo {author} {\bibfnamefont {V.}~\bibnamefont {Sharma}},
  \bibinfo {author} {\bibfnamefont {A.~J.}\ \bibnamefont {Biacchi}}, \bibinfo
  {author} {\bibfnamefont {A.~R.}\ \bibnamefont {Hight~Walker}},\ and\ \bibinfo
  {author} {\bibfnamefont {F.}~\bibnamefont {Tavazza}},\ }\bibfield  {title}
  {\bibinfo {title} {High-throughput density functional perturbation theory and
  machine learning predictions of infrared, piezoelectric, and dielectric
  responses},\ }\href {https://doi.org/10.1038/s41524-020-0337-2} {\bibfield
  {journal} {\bibinfo  {journal} {npj Comput. Mater.}\ }\textbf {\bibinfo
  {volume} {6}},\ \bibinfo {pages} {64} (\bibinfo {year} {2020})}\BibitemShut
  {NoStop}%
\bibitem [{\citenamefont {Fang}\ \emph {et~al.}(2024)\citenamefont {Fang},
  \citenamefont {Geiger}, \citenamefont {Checkelsky},\ and\ \citenamefont
  {Smidt}}]{fang2024arxiv}%
  \BibitemOpen
  \bibfield  {author} {\bibinfo {author} {\bibfnamefont {S.}~\bibnamefont
  {Fang}}, \bibinfo {author} {\bibfnamefont {M.}~\bibnamefont {Geiger}},
  \bibinfo {author} {\bibfnamefont {J.~G.}\ \bibnamefont {Checkelsky}},\ and\
  \bibinfo {author} {\bibfnamefont {T.}~\bibnamefont {Smidt}},\ }\href@noop {}
  {\bibinfo {title} {Phonon predictions with {E}(3)-equivariant graph neural
  networks}} (\bibinfo {year} {2024}),\ \Eprint
  {https://arxiv.org/abs/2403.11347} {arXiv:2403.11347} \BibitemShut {NoStop}%
\bibitem [{\citenamefont {Schmiedmayer}\ and\ \citenamefont
  {Kresse}(2024)}]{schmiedmayer2024arxiv}%
  \BibitemOpen
  \bibfield  {author} {\bibinfo {author} {\bibfnamefont {B.}~\bibnamefont
  {Schmiedmayer}}\ and\ \bibinfo {author} {\bibfnamefont {G.}~\bibnamefont
  {Kresse}},\ }\bibfield  {title} {\bibinfo {title} {Derivative learning of
  tensorial quantities--predicting finite temperature infrared spectra from
  first principles},\ }\href {https://arxiv.org/abs/2404.19674} {\bibfield
  {journal} {\bibinfo  {journal} {arXiv:2404.19674}\ } (\bibinfo {year}
  {2024})}\BibitemShut {NoStop}%
\bibitem [{\citenamefont {Wilkins}\ \emph {et~al.}(2019)\citenamefont
  {Wilkins}, \citenamefont {Grisafi}, \citenamefont {Yang}, \citenamefont
  {Lao}, \citenamefont {DiStasio},\ and\ \citenamefont
  {Ceriotti}}]{wilkins2019PNAS}%
  \BibitemOpen
  \bibfield  {author} {\bibinfo {author} {\bibfnamefont {D.~M.}\ \bibnamefont
  {Wilkins}}, \bibinfo {author} {\bibfnamefont {A.}~\bibnamefont {Grisafi}},
  \bibinfo {author} {\bibfnamefont {Y.}~\bibnamefont {Yang}}, \bibinfo {author}
  {\bibfnamefont {K.~U.}\ \bibnamefont {Lao}}, \bibinfo {author} {\bibfnamefont
  {R.~A.}\ \bibnamefont {DiStasio}},\ and\ \bibinfo {author} {\bibfnamefont
  {M.}~\bibnamefont {Ceriotti}},\ }\bibfield  {title} {\bibinfo {title}
  {Accurate molecular polarizabilities with coupled cluster theory and machine
  learning},\ }\href {https://doi.org/10.1073/pnas.1816132116} {\bibfield
  {journal} {\bibinfo  {journal} {Proc. Natl. Acad. Sci.}\ }\textbf {\bibinfo
  {volume} {116}},\ \bibinfo {pages} {3401} (\bibinfo {year}
  {2019})}\BibitemShut {NoStop}%
\bibitem [{\citenamefont {Sommers}\ \emph {et~al.}(2020)\citenamefont
  {Sommers}, \citenamefont {Calegari~Andrade}, \citenamefont {Zhang},
  \citenamefont {Wang},\ and\ \citenamefont {Car}}]{sommers2020PCCP}%
  \BibitemOpen
  \bibfield  {author} {\bibinfo {author} {\bibfnamefont {G.~M.}\ \bibnamefont
  {Sommers}}, \bibinfo {author} {\bibfnamefont {M.~F.}\ \bibnamefont
  {Calegari~Andrade}}, \bibinfo {author} {\bibfnamefont {L.}~\bibnamefont
  {Zhang}}, \bibinfo {author} {\bibfnamefont {H.}~\bibnamefont {Wang}},\ and\
  \bibinfo {author} {\bibfnamefont {R.}~\bibnamefont {Car}},\ }\bibfield
  {title} {\bibinfo {title} {{R}aman spectrum and polarizability of liquid
  water from deep neural networks},\ }\href
  {https://doi.org/10.1039/D0CP01893G} {\bibfield  {journal} {\bibinfo
  {journal} {Phys. Chem. Chem. Phys.}\ }\textbf {\bibinfo {volume} {22}},\
  \bibinfo {pages} {10592} (\bibinfo {year} {2020})}\BibitemShut {NoStop}%
\bibitem [{\citenamefont {Takahashi}\ \emph {et~al.}(2020)\citenamefont
  {Takahashi}, \citenamefont {Kumagai}, \citenamefont {Miyamoto}, \citenamefont
  {Mochizuki},\ and\ \citenamefont {Oba}}]{takahashi2020PRM}%
  \BibitemOpen
  \bibfield  {author} {\bibinfo {author} {\bibfnamefont {A.}~\bibnamefont
  {Takahashi}}, \bibinfo {author} {\bibfnamefont {Y.}~\bibnamefont {Kumagai}},
  \bibinfo {author} {\bibfnamefont {J.}~\bibnamefont {Miyamoto}}, \bibinfo
  {author} {\bibfnamefont {Y.}~\bibnamefont {Mochizuki}},\ and\ \bibinfo
  {author} {\bibfnamefont {F.}~\bibnamefont {Oba}},\ }\bibfield  {title}
  {\bibinfo {title} {Machine learning models for predicting the dielectric
  constants of oxides based on high-throughput first-principles calculations},\
  }\href {https://doi.org/10.1103/PhysRevMaterials.4.103801} {\bibfield
  {journal} {\bibinfo  {journal} {Phys. Rev. Mater.}\ }\textbf {\bibinfo
  {volume} {4}},\ \bibinfo {pages} {103801} (\bibinfo {year}
  {2020})}\BibitemShut {NoStop}%
\bibitem [{\citenamefont {Kapil}\ \emph {et~al.}(2024)\citenamefont {Kapil},
  \citenamefont {Kovács}, \citenamefont {Csányi},\ and\ \citenamefont
  {Michaelides}}]{kapil2023FD}%
  \BibitemOpen
  \bibfield  {author} {\bibinfo {author} {\bibfnamefont {V.}~\bibnamefont
  {Kapil}}, \bibinfo {author} {\bibfnamefont {D.~P.}\ \bibnamefont {Kovács}},
  \bibinfo {author} {\bibfnamefont {G.}~\bibnamefont {Csányi}},\ and\ \bibinfo
  {author} {\bibfnamefont {A.}~\bibnamefont {Michaelides}},\ }\bibfield
  {title} {\bibinfo {title} {First-principles spectroscopy of aqueous
  interfaces using machine-learned electronic and quantum nuclear effects},\
  }\href {https://doi.org/10.1039/D3FD00113J} {\bibfield  {journal} {\bibinfo
  {journal} {Faraday Discuss.}\ }\textbf {\bibinfo {volume} {249}},\ \bibinfo
  {pages} {50} (\bibinfo {year} {2024})}\BibitemShut {NoStop}%
\bibitem [{\citenamefont {Berger}\ and\ \citenamefont
  {Komsa}(2024)}]{berger2024PRM}%
  \BibitemOpen
  \bibfield  {author} {\bibinfo {author} {\bibfnamefont {E.}~\bibnamefont
  {Berger}}\ and\ \bibinfo {author} {\bibfnamefont {H.-P.}\ \bibnamefont
  {Komsa}},\ }\bibfield  {title} {\bibinfo {title} {Polarizability models for
  simulations of finite temperature raman spectra from machine learning
  molecular dynamics},\ }\href
  {https://doi.org/10.1103/PhysRevMaterials.8.043802} {\bibfield  {journal}
  {\bibinfo  {journal} {Phys. Rev. Mater.}\ }\textbf {\bibinfo {volume} {8}},\
  \bibinfo {pages} {043802} (\bibinfo {year} {2024})}\BibitemShut {NoStop}%
\bibitem [{\citenamefont {Raissi}\ \emph {et~al.}(2017)\citenamefont {Raissi},
  \citenamefont {Perdikaris},\ and\ \citenamefont
  {Karniadakis}}]{raissi2017arxiv}%
  \BibitemOpen
  \bibfield  {author} {\bibinfo {author} {\bibfnamefont {M.}~\bibnamefont
  {Raissi}}, \bibinfo {author} {\bibfnamefont {P.}~\bibnamefont {Perdikaris}},\
  and\ \bibinfo {author} {\bibfnamefont {G.~E.}\ \bibnamefont {Karniadakis}},\
  }\bibfield  {title} {\bibinfo {title} {Physics informed deep learning {(Part
  I)}: Data-driven solutions of nonlinear partial differential equations},\
  }\href {https://arxiv.org/abs/1711.10561} {\bibfield  {journal} {\bibinfo
  {journal} {arXiv:1711.10561}\ } (\bibinfo {year} {2017})}\BibitemShut
  {NoStop}%
\bibitem [{\citenamefont {Czarnecki}\ \emph {et~al.}(2017)\citenamefont
  {Czarnecki}, \citenamefont {Osindero}, \citenamefont {Jaderberg},
  \citenamefont {Świrszcz},\ and\ \citenamefont
  {Pascanu}}]{czarnecki2017arxiv}%
  \BibitemOpen
  \bibfield  {author} {\bibinfo {author} {\bibfnamefont {W.~M.}\ \bibnamefont
  {Czarnecki}}, \bibinfo {author} {\bibfnamefont {S.}~\bibnamefont {Osindero}},
  \bibinfo {author} {\bibfnamefont {M.}~\bibnamefont {Jaderberg}}, \bibinfo
  {author} {\bibfnamefont {G.}~\bibnamefont {Świrszcz}},\ and\ \bibinfo
  {author} {\bibfnamefont {R.}~\bibnamefont {Pascanu}},\ }\bibfield  {title}
  {\bibinfo {title} {Sobolev training for neural networks},\ }\href
  {https://arxiv.org/abs/1706.04859} {\bibfield  {journal} {\bibinfo  {journal}
  {arXiv:1706.04859}\ } (\bibinfo {year} {2017})}\BibitemShut {NoStop}%
\bibitem [{\citenamefont {Vandermause}\ \emph {et~al.}(2020)\citenamefont
  {Vandermause}, \citenamefont {Torrisi}, \citenamefont {Batzner},
  \citenamefont {Xie}, \citenamefont {Sun}, \citenamefont {Kolpak},\ and\
  \citenamefont {Kozinsky}}]{vandermause2020npj}%
  \BibitemOpen
  \bibfield  {author} {\bibinfo {author} {\bibfnamefont {J.}~\bibnamefont
  {Vandermause}}, \bibinfo {author} {\bibfnamefont {S.~B.}\ \bibnamefont
  {Torrisi}}, \bibinfo {author} {\bibfnamefont {S.}~\bibnamefont {Batzner}},
  \bibinfo {author} {\bibfnamefont {Y.}~\bibnamefont {Xie}}, \bibinfo {author}
  {\bibfnamefont {L.}~\bibnamefont {Sun}}, \bibinfo {author} {\bibfnamefont
  {A.~M.}\ \bibnamefont {Kolpak}},\ and\ \bibinfo {author} {\bibfnamefont
  {B.}~\bibnamefont {Kozinsky}},\ }\bibfield  {title} {\bibinfo {title}
  {On-the-fly active learning of interpretable bayesian force fields for
  atomistic rare events},\ }\href {https://doi.org/10.1038/s41524-020-0283-z}
  {\bibfield  {journal} {\bibinfo  {journal} {npj Comput. Mater.}\ }\textbf
  {\bibinfo {volume} {6}},\ \bibinfo {pages} {20} (\bibinfo {year}
  {2020})}\BibitemShut {NoStop}%
\bibitem [{\citenamefont {Duschatko}\ \emph {et~al.}(2024)\citenamefont
  {Duschatko}, \citenamefont {Fu}, \citenamefont {Owen}, \citenamefont {Xie},
  \citenamefont {Musaelian}, \citenamefont {Jaakkola},\ and\ \citenamefont
  {Kozinsky}}]{duschatko2024arxiv}%
  \BibitemOpen
  \bibfield  {author} {\bibinfo {author} {\bibfnamefont {B.~R.}\ \bibnamefont
  {Duschatko}}, \bibinfo {author} {\bibfnamefont {X.}~\bibnamefont {Fu}},
  \bibinfo {author} {\bibfnamefont {C.}~\bibnamefont {Owen}}, \bibinfo {author}
  {\bibfnamefont {Y.}~\bibnamefont {Xie}}, \bibinfo {author} {\bibfnamefont
  {A.}~\bibnamefont {Musaelian}}, \bibinfo {author} {\bibfnamefont
  {T.}~\bibnamefont {Jaakkola}},\ and\ \bibinfo {author} {\bibfnamefont
  {B.}~\bibnamefont {Kozinsky}},\ }\bibfield  {title} {\bibinfo {title}
  {Thermodynamically informed multimodal learning of high-dimensional free
  energy models in molecular coarse graining},\ }\href
  {https://arxiv.org/abs/2405.19386} {\bibfield  {journal} {\bibinfo  {journal}
  {arXiv:2405.19386}\ } (\bibinfo {year} {2024})}\BibitemShut {NoStop}%
\bibitem [{\citenamefont {Thompson}\ \emph {et~al.}(2022)\citenamefont
  {Thompson}, \citenamefont {Aktulga}, \citenamefont {Berger}, \citenamefont
  {Bolintineanu}, \citenamefont {Brown}, \citenamefont {Crozier}, \citenamefont
  {{in 't Veld}}, \citenamefont {Kohlmeyer}, \citenamefont {Moore},
  \citenamefont {Nguyen}, \citenamefont {Shan}, \citenamefont {Stevens},
  \citenamefont {Tranchida}, \citenamefont {Trott},\ and\ \citenamefont
  {Plimpton}}]{thompson2022CPC}%
  \BibitemOpen
  \bibfield  {author} {\bibinfo {author} {\bibfnamefont {A.~P.}\ \bibnamefont
  {Thompson}}, \bibinfo {author} {\bibfnamefont {H.~M.}\ \bibnamefont
  {Aktulga}}, \bibinfo {author} {\bibfnamefont {R.}~\bibnamefont {Berger}},
  \bibinfo {author} {\bibfnamefont {D.~S.}\ \bibnamefont {Bolintineanu}},
  \bibinfo {author} {\bibfnamefont {W.~M.}\ \bibnamefont {Brown}}, \bibinfo
  {author} {\bibfnamefont {P.~S.}\ \bibnamefont {Crozier}}, \bibinfo {author}
  {\bibfnamefont {P.~J.}\ \bibnamefont {{in 't Veld}}}, \bibinfo {author}
  {\bibfnamefont {A.}~\bibnamefont {Kohlmeyer}}, \bibinfo {author}
  {\bibfnamefont {S.~G.}\ \bibnamefont {Moore}}, \bibinfo {author}
  {\bibfnamefont {T.~D.}\ \bibnamefont {Nguyen}}, \bibinfo {author}
  {\bibfnamefont {R.}~\bibnamefont {Shan}}, \bibinfo {author} {\bibfnamefont
  {M.~J.}\ \bibnamefont {Stevens}}, \bibinfo {author} {\bibfnamefont
  {J.}~\bibnamefont {Tranchida}}, \bibinfo {author} {\bibfnamefont
  {C.}~\bibnamefont {Trott}},\ and\ \bibinfo {author} {\bibfnamefont {S.~J.}\
  \bibnamefont {Plimpton}},\ }\bibfield  {title} {\bibinfo {title} {{LAMMPS} -
  a flexible simulation tool for particle-based materials modeling at the
  atomic, meso, and continuum scales},\ }\href
  {https://doi.org/https://doi.org/10.1016/j.cpc.2021.108171} {\bibfield
  {journal} {\bibinfo  {journal} {Comput. Phys. Commun.}\ }\textbf {\bibinfo
  {volume} {271}},\ \bibinfo {pages} {108171} (\bibinfo {year}
  {2022})}\BibitemShut {NoStop}%
\bibitem [{\citenamefont {Giacomazzi}\ \emph {et~al.}(2009)\citenamefont
  {Giacomazzi}, \citenamefont {Umari},\ and\ \citenamefont
  {Pasquarello}}]{giacomazzi2009PRB}%
  \BibitemOpen
  \bibfield  {author} {\bibinfo {author} {\bibfnamefont {L.}~\bibnamefont
  {Giacomazzi}}, \bibinfo {author} {\bibfnamefont {P.}~\bibnamefont {Umari}},\
  and\ \bibinfo {author} {\bibfnamefont {A.}~\bibnamefont {Pasquarello}},\
  }\bibfield  {title} {\bibinfo {title} {Medium-range structure of vitreous
  {SiO$_2$} obtained through first-principles investigation of vibrational
  spectra},\ }\href {https://doi.org/10.1103/PhysRevB.79.064202} {\bibfield
  {journal} {\bibinfo  {journal} {Phys. Rev. B}\ }\textbf {\bibinfo {volume}
  {79}},\ \bibinfo {pages} {064202} (\bibinfo {year} {2009})}\BibitemShut
  {NoStop}%
\bibitem [{\citenamefont {Akbarian}\ \emph {et~al.}(2019)\citenamefont
  {Akbarian}, \citenamefont {Yilmaz}, \citenamefont {Cao}, \citenamefont
  {Ganesh}, \citenamefont {Dabo}, \citenamefont {Munro}, \citenamefont
  {Van~Ginhoven},\ and\ \citenamefont {van Duin}}]{akbarian2019PCCP}%
  \BibitemOpen
  \bibfield  {author} {\bibinfo {author} {\bibfnamefont {D.}~\bibnamefont
  {Akbarian}}, \bibinfo {author} {\bibfnamefont {D.~E.}\ \bibnamefont
  {Yilmaz}}, \bibinfo {author} {\bibfnamefont {Y.}~\bibnamefont {Cao}},
  \bibinfo {author} {\bibfnamefont {P.}~\bibnamefont {Ganesh}}, \bibinfo
  {author} {\bibfnamefont {I.}~\bibnamefont {Dabo}}, \bibinfo {author}
  {\bibfnamefont {J.}~\bibnamefont {Munro}}, \bibinfo {author} {\bibfnamefont
  {R.}~\bibnamefont {Van~Ginhoven}},\ and\ \bibinfo {author} {\bibfnamefont
  {A.~C.~T.}\ \bibnamefont {van Duin}},\ }\bibfield  {title} {\bibinfo {title}
  {Understanding the influence of defects and surface chemistry on
  ferroelectric switching: a reaxff investigation of batio3},\ }\href
  {https://doi.org/10.1039/C9CP02955A} {\bibfield  {journal} {\bibinfo
  {journal} {Phys. Chem. Chem. Phys.}\ }\textbf {\bibinfo {volume} {21}},\
  \bibinfo {pages} {18240} (\bibinfo {year} {2019})}\BibitemShut {NoStop}%
\bibitem [{\citenamefont {Deguchi}\ \emph {et~al.}(2023)\citenamefont
  {Deguchi}, \citenamefont {Kobayashi}, \citenamefont {Azuma}, \citenamefont
  {Ogata}, \citenamefont {Uranagase},\ and\ \citenamefont
  {Spreafico}}]{deguchi2023PSS}%
  \BibitemOpen
  \bibfield  {author} {\bibinfo {author} {\bibfnamefont {G.}~\bibnamefont
  {Deguchi}}, \bibinfo {author} {\bibfnamefont {R.}~\bibnamefont {Kobayashi}},
  \bibinfo {author} {\bibfnamefont {H.}~\bibnamefont {Azuma}}, \bibinfo
  {author} {\bibfnamefont {S.}~\bibnamefont {Ogata}}, \bibinfo {author}
  {\bibfnamefont {M.}~\bibnamefont {Uranagase}},\ and\ \bibinfo {author}
  {\bibfnamefont {S.}~\bibnamefont {Spreafico}},\ }\bibfield  {title} {\bibinfo
  {title} {Asymmetric domain nucleation from dislocation core in barium
  titanate: Molecular dynamics simulation using machine-learning potential
  through active learning},\ }\href
  {https://doi.org/https://doi.org/10.1002/pssr.202300292} {\bibfield
  {journal} {\bibinfo  {journal} {Phys. Status Solidi}\ ,\ \bibinfo {pages}
  {2300292}} (\bibinfo {year} {2023})}\BibitemShut {NoStop}%
\bibitem [{\citenamefont {Loose}\ \emph {et~al.}(2023)\citenamefont {Loose},
  \citenamefont {Sahrmann}, \citenamefont {Qu},\ and\ \citenamefont
  {Voth}}]{loose2023JPCB}%
  \BibitemOpen
  \bibfield  {author} {\bibinfo {author} {\bibfnamefont {T.~D.}\ \bibnamefont
  {Loose}}, \bibinfo {author} {\bibfnamefont {P.~G.}\ \bibnamefont {Sahrmann}},
  \bibinfo {author} {\bibfnamefont {T.~S.}\ \bibnamefont {Qu}},\ and\ \bibinfo
  {author} {\bibfnamefont {G.~A.}\ \bibnamefont {Voth}},\ }\bibfield  {title}
  {\bibinfo {title} {Coarse-graining with equivariant neural networks: A path
  toward accurate and data-efficient models},\ }\href
  {https://doi.org/10.1021/acs.jpcb.3c05928} {\bibfield  {journal} {\bibinfo
  {journal} {J. Phys. Chem. B}\ }\textbf {\bibinfo {volume} {127}},\ \bibinfo
  {pages} {10564} (\bibinfo {year} {2023})}\BibitemShut {NoStop}%
\bibitem [{\citenamefont {Fu}\ \emph {et~al.}(2023)\citenamefont {Fu},
  \citenamefont {Wu}, \citenamefont {Wang}, \citenamefont {Xie}, \citenamefont
  {Keten}, \citenamefont {Gomez-Bombarelli},\ and\ \citenamefont
  {Jaakkola}}]{fu2023arxiv}%
  \BibitemOpen
  \bibfield  {author} {\bibinfo {author} {\bibfnamefont {X.}~\bibnamefont
  {Fu}}, \bibinfo {author} {\bibfnamefont {Z.}~\bibnamefont {Wu}}, \bibinfo
  {author} {\bibfnamefont {W.}~\bibnamefont {Wang}}, \bibinfo {author}
  {\bibfnamefont {T.}~\bibnamefont {Xie}}, \bibinfo {author} {\bibfnamefont
  {S.}~\bibnamefont {Keten}}, \bibinfo {author} {\bibfnamefont
  {R.}~\bibnamefont {Gomez-Bombarelli}},\ and\ \bibinfo {author} {\bibfnamefont
  {T.}~\bibnamefont {Jaakkola}},\ }\href@noop {} {\bibinfo {title} {Forces are
  not enough: Benchmark and critical evaluation for machine learning force
  fields with molecular simulations}} (\bibinfo {year} {2023}),\ \Eprint
  {https://arxiv.org/abs/2210.07237} {arXiv:2210.07237} \BibitemShut {NoStop}%
\bibitem [{\citenamefont {Maxson}\ and\ \citenamefont
  {Szilvasi}(2024)}]{maxson2024arxiv}%
  \BibitemOpen
  \bibfield  {author} {\bibinfo {author} {\bibfnamefont {T.}~\bibnamefont
  {Maxson}}\ and\ \bibinfo {author} {\bibfnamefont {T.}~\bibnamefont
  {Szilvasi}},\ }\href@noop {} {\bibinfo {title} {Transferable water potentials
  using equivariant neural networks}} (\bibinfo {year} {2024}),\ \Eprint
  {https://arxiv.org/abs/2402.16204} {arXiv:2402.16204} \BibitemShut {NoStop}%
\bibitem [{\citenamefont {Broughton}\ \emph {et~al.}(1997)\citenamefont
  {Broughton}, \citenamefont {Meli}, \citenamefont {Vashishta},\ and\
  \citenamefont {Kalia}}]{broughton1997PRB}%
  \BibitemOpen
  \bibfield  {author} {\bibinfo {author} {\bibfnamefont {J.~Q.}\ \bibnamefont
  {Broughton}}, \bibinfo {author} {\bibfnamefont {C.~A.}\ \bibnamefont {Meli}},
  \bibinfo {author} {\bibfnamefont {P.}~\bibnamefont {Vashishta}},\ and\
  \bibinfo {author} {\bibfnamefont {R.~K.}\ \bibnamefont {Kalia}},\ }\bibfield
  {title} {\bibinfo {title} {Direct atomistic simulation of quartz crystal
  oscillators: Bulk properties and nanoscale devices},\ }\href
  {https://doi.org/10.1103/PhysRevB.56.611} {\bibfield  {journal} {\bibinfo
  {journal} {Phys. Rev. B}\ }\textbf {\bibinfo {volume} {56}},\ \bibinfo
  {pages} {611} (\bibinfo {year} {1997})}\BibitemShut {NoStop}%
\bibitem [{\citenamefont {Giannozzi}\ \emph {et~al.}(2009)\citenamefont
  {Giannozzi}, \citenamefont {Baroni}, \citenamefont {Bonini}, \citenamefont
  {Calandra}, \citenamefont {Car}, \citenamefont {Cavazzoni}, \citenamefont
  {Ceresoli}, \citenamefont {Chiarotti}, \citenamefont {Cococcioni},
  \citenamefont {Dabo}, \citenamefont {Corso}, \citenamefont {de~Gironcoli},
  \citenamefont {Fabris}, \citenamefont {Fratesi}, \citenamefont {Gebauer},
  \citenamefont {Gerstmann}, \citenamefont {Gougoussis}, \citenamefont
  {Kokalj}, \citenamefont {Lazzeri}, \citenamefont {Martin-Samos},
  \citenamefont {Marzari}, \citenamefont {Mauri}, \citenamefont {Mazzarello},
  \citenamefont {Paolini}, \citenamefont {Pasquarello}, \citenamefont
  {Paulatto}, \citenamefont {Sbraccia}, \citenamefont {Scandolo}, \citenamefont
  {Sclauzero}, \citenamefont {Seitsonen}, \citenamefont {Smogunov},
  \citenamefont {Umari},\ and\ \citenamefont
  {Wentzcovitch}}]{giannozzi2009JPCM}%
  \BibitemOpen
  \bibfield  {author} {\bibinfo {author} {\bibfnamefont {P.}~\bibnamefont
  {Giannozzi}}, \bibinfo {author} {\bibfnamefont {S.}~\bibnamefont {Baroni}},
  \bibinfo {author} {\bibfnamefont {N.}~\bibnamefont {Bonini}}, \bibinfo
  {author} {\bibfnamefont {M.}~\bibnamefont {Calandra}}, \bibinfo {author}
  {\bibfnamefont {R.}~\bibnamefont {Car}}, \bibinfo {author} {\bibfnamefont
  {C.}~\bibnamefont {Cavazzoni}}, \bibinfo {author} {\bibfnamefont
  {D.}~\bibnamefont {Ceresoli}}, \bibinfo {author} {\bibfnamefont {G.~L.}\
  \bibnamefont {Chiarotti}}, \bibinfo {author} {\bibfnamefont {M.}~\bibnamefont
  {Cococcioni}}, \bibinfo {author} {\bibfnamefont {I.}~\bibnamefont {Dabo}},
  \bibinfo {author} {\bibfnamefont {A.~D.}\ \bibnamefont {Corso}}, \bibinfo
  {author} {\bibfnamefont {S.}~\bibnamefont {de~Gironcoli}}, \bibinfo {author}
  {\bibfnamefont {S.}~\bibnamefont {Fabris}}, \bibinfo {author} {\bibfnamefont
  {G.}~\bibnamefont {Fratesi}}, \bibinfo {author} {\bibfnamefont
  {R.}~\bibnamefont {Gebauer}}, \bibinfo {author} {\bibfnamefont
  {U.}~\bibnamefont {Gerstmann}}, \bibinfo {author} {\bibfnamefont
  {C.}~\bibnamefont {Gougoussis}}, \bibinfo {author} {\bibfnamefont
  {A.}~\bibnamefont {Kokalj}}, \bibinfo {author} {\bibfnamefont
  {M.}~\bibnamefont {Lazzeri}}, \bibinfo {author} {\bibfnamefont
  {L.}~\bibnamefont {Martin-Samos}}, \bibinfo {author} {\bibfnamefont
  {N.}~\bibnamefont {Marzari}}, \bibinfo {author} {\bibfnamefont
  {F.}~\bibnamefont {Mauri}}, \bibinfo {author} {\bibfnamefont
  {R.}~\bibnamefont {Mazzarello}}, \bibinfo {author} {\bibfnamefont
  {S.}~\bibnamefont {Paolini}}, \bibinfo {author} {\bibfnamefont
  {A.}~\bibnamefont {Pasquarello}}, \bibinfo {author} {\bibfnamefont
  {L.}~\bibnamefont {Paulatto}}, \bibinfo {author} {\bibfnamefont
  {C.}~\bibnamefont {Sbraccia}}, \bibinfo {author} {\bibfnamefont
  {S.}~\bibnamefont {Scandolo}}, \bibinfo {author} {\bibfnamefont
  {G.}~\bibnamefont {Sclauzero}}, \bibinfo {author} {\bibfnamefont {A.~P.}\
  \bibnamefont {Seitsonen}}, \bibinfo {author} {\bibfnamefont {A.}~\bibnamefont
  {Smogunov}}, \bibinfo {author} {\bibfnamefont {P.}~\bibnamefont {Umari}},\
  and\ \bibinfo {author} {\bibfnamefont {R.~M.}\ \bibnamefont {Wentzcovitch}},\
  }\bibfield  {title} {\bibinfo {title} {{QUANTUM ESPRESSO}: a modular and
  open-source software project for quantum simulations of materials},\ }\href
  {https://doi.org/10.1088/0953-8984/21/39/395502} {\bibfield  {journal}
  {\bibinfo  {journal} {J. Phys.: Condens. Matter}\ }\textbf {\bibinfo {volume}
  {21}},\ \bibinfo {pages} {395502} (\bibinfo {year} {2009})}\BibitemShut
  {NoStop}%
\bibitem [{\citenamefont {Marsalek}\ and\ \citenamefont
  {Markland}(2017)}]{marsalek2017PRL}%
  \BibitemOpen
  \bibfield  {author} {\bibinfo {author} {\bibfnamefont {O.}~\bibnamefont
  {Marsalek}}\ and\ \bibinfo {author} {\bibfnamefont {T.~E.}\ \bibnamefont
  {Markland}},\ }\bibfield  {title} {\bibinfo {title} {Quantum dynamics and
  spectroscopy of ab initio liquid water: The interplay of nuclear and
  electronic quantum effects},\ }\href
  {https://doi.org/10.1021/acs.jpclett.7b00391} {\bibfield  {journal} {\bibinfo
   {journal} {J. Phys. Chem. Lett.}\ }\textbf {\bibinfo {volume} {8}},\
  \bibinfo {pages} {1545} (\bibinfo {year} {2017})}\BibitemShut {NoStop}%
\end{thebibliography}%

\newpage 

\vspace{0.3cm} \noindent {\large \textbf{Methods}} \vspace{0.1cm}

\noindent \textbf{Theory}\\
Our goal is to learn the generalized potential energy $U$ of a system that depends on atomic coordinates $r_{i\nu}$ and on a set of parameters $\lambda_{\mu}$, where $i$ is the atom index, and $\nu$ and $\mu$ are Cartesian indexes. We start by expanding $U$ in a Taylor series, namely
\begin{align}
& U(r_{i\nu}+\delta r_{i\nu}, \lambda_\mu+\delta \lambda_\mu) \nonumber \\
& \ = U(r_{i\nu}, \lambda_\mu) + \frac{\partial U}{\partial r_{i\nu}} \delta r_{i\nu} + \frac{\partial U}{\partial \lambda_\mu} \delta \lambda_\mu \nonumber \\
& \quad 
+\frac{1}{2} \frac{\partial^2 U}{\partial \lambda_\mu \partial \lambda_{\mu'}} \delta \lambda_\mu \delta \lambda_{\mu'} + \frac{1}{2} \frac{\partial^2 U}{\partial r_{i\nu} \partial \lambda_{\mu}} \delta r_{i\nu}\delta \lambda_\mu+ \dots\label{eq:e_expansion}
\end{align}
where we employ the summation convention. The parameters $\lambda_\mu$ can include lattice vectors, volume, electric field, magnetic field, electrostatic or chemical potential. The corresponding conjugate properties $\partial U/\partial \lambda_\mu$ include stress, pressure, polarization, magnetization, electronic charge or particle number.

In the case of a uniform electric field $\mathbf E$, the electric enthalpy functional is defined as \cite{nunes2001PRB,umari2002PRL} 
\begin{equation}
    U = U^0 - \mathbf{E}\cdot \mathbf P, \label{eq:enthalpy}
\end{equation}
where $U^0$ is the energy in the absence of an electric field, and $\mathbf P$ the polarization.  The polarization is related to the electric enthalpy through the following differential relation,
\begin{equation}
    P_\mu = - \frac{\partial U}{\partial E_\mu}. \label{eq:pol_E}
\end{equation}
In the modern theory of polarization, $\mathbf P$ is a multivalued quantity defined modulo the quantum of polarization $\Delta \mathbf P = e \mathbf R$, where $\mathbf R$ is a lattice vector. In the limit of a weak electric field, the derivatives of the polarization with respect to the atomic displacements define the Born charges,
\begin{equation}
    Z^*_{i \mu \nu} = \left.\frac{1}{e} \frac{\partial P_\mu}{\partial r_{i \nu}}\right|_{E_\mu = 0}.\label{eq:born_charge}
\end{equation}
Born charges obey the acoustic sum rule, namely $\sum_i Z^*_{i \mu\nu} = 0$. 
This reflects the translation invariance of polarization and the charge neutrality of the system. The derivative of the polarization with respect to the electric field defines the polarizability,
\begin{equation}
    \alpha_{\mu\nu} = \frac{\partial P_\mu}{\partial E_\nu}. \label{eq:polarizability}
\end{equation}

\vspace{0.3cm}
\noindent \textbf{Neural network architecture}\\
We use the differential relations in equations \eqref{eq:pol_E}-\eqref{eq:polarizability} to implement the learning approach for polarization, Born charges and polarizability in addition to electric enthalpy, forces, and stress within a unified framework. This is implemented in the Allegro code \cite{musaelian2023NC} as follows. First, we include the electric field as an input of the network along with the atomic positions. In particular, the spherical harmonics embedding for the electric field and interatomic displacements are concatenated and treated on the same footing as geometric vector quantities within the neural network architecture \cite{musaelian2023NC}. Thanks to the modularity of the Allegro code, this incorporation is achieved without significant modifications of its core architecture. Then, we determine the polarization by differentiating the output of the model, i.e.\ the electric enthalpy, with respect to the electric field at its zero value. This procedure is analogous to how the atomic forces are obtained as derivatives of the electric enthalpy with respect to the atomic positions (cf.\ Fig.\ \ref{fig:model}a). Then, following equations \eqref{eq:born_charge} and \eqref{eq:polarizability}, we derive Born charges and polarizability by differentiating the polarization with respect to atomic positions and electric field, respectively. Since Born charges are per-atom quantities, they provide a large amount of information for learning the polarization, thus increasing data efficiency. 

Polarization, Born charges and polarizability are learned along with the electric enthalpy, atomic forces, and stress by adding the following extra contribution to the conventional force field loss function:
\begin{align}
    \Delta \mathcal L & = \frac{\lambda_P}{3N} \sum_{\mu=1}^3 \left| \left(-\frac{\partial\hat{U}}{\partial E_\mu} - P_\mu \right) \text{mod} \ \Delta P_\mu \right| \nonumber \\
    & \qquad  + \frac{\lambda_{Z}}{9N} \sum_{i=1}^N \sum_{\mu=1}^3 \sum_{\nu=1}^3 \left| -\frac{\partial^2\hat{U}}{\partial E_\mu \partial r_{i\nu}}- Z^*_{i\mu\nu}\right| \nonumber \\
    & \qquad  + \frac{\lambda_{\alpha}}{9N} \sum_{\mu=1}^3 \sum_{\nu=1}^3 \left| -\frac{\partial^2\hat{U}}{\partial E_\mu \partial E_\nu } - \alpha_{\mu\nu} \right|, 
    \label{eq:loss}
\end{align}
where $\lambda_{P}$, $\lambda_{Z}$, and $\lambda_{\alpha}$ are the loss weights, $N$ is the number of atoms, $\hat{U}$ the predicted electric enthalpy, and $\Delta \mathbf P$ the quantum of polarization. In equation \eqref{eq:loss}, the loss contributions of extensive quantities, namely polarization and polarizability, are normalized by the number of atoms. Moreover, the loss contribution related to the polarization is computed with the minimum-image convention to account for the multivalued nature of polarization in the Berry phase theory. This overcomes issues related to training on polarization values belonging to multiple branches of the Berry phase \cite{schmiedmayer2024arxiv}, and avoids the use of pre-training strategies based on folding polarization values within the same Berry phase branch \cite{gigli2022npj}. In addition, we remark that the training on polarization values is not affected by the choice of the origin of the simulation cell. Indeed, the DFT labels of polarization are independent of the origin due to the charge neutrality of the system \cite{resta1994RMP}, and our ML model is translation invariant since its representation is based on interatomic displacements rather than atomic positions. Our ML models for $\alpha$-\ce{SiO2} and \ce{BaTiO3} were  exhaustively tested over various hyperparameters, including cutoff radius, maximum order of spherical harmonics, loss coefficients, number of tensor features, and network architecture. We find that generally a maximum order of spherical harmonics of 3 or greater improves the model, especially for simpler network architectures. Complete computational details are provided in the Supplementary Information (SI).

We note that the multivalued nature of polarization in DFT poses extra challenges, as one needs to consistently fold polarization values from different Berry phase branches into a single branch. In contrast, our ML model predicts polarization values that vary smoothly as a function of the electric field. This is because our model is based on neural networks, which are suited for modeling continuous functions. Additionally, the learning of polarization is primarily driven by the learning of Born charges, which are related to polarization through a derivative relation and for which effects due to the Berry phase branch vanish.

Our method presents several practical advantages. First, it is based on equivariant local representations of the atomic environment, eliminating the need for message passing present in other graph neural network models. This is crucial for performing large-scale MPI-parallelized MD simulations, as the receptive field of message-passing neural networks can grow excessively, constraining parallel computation and the scale of MD simulations \cite{musaelian2023NC}. Second, the model predicts a scalar, i.e.\ the electric enthalpy, and derives therefrom all response quantities for inference and training through automatic differentiation. Predicting a scalar requires a minimal number of tensorial paths and is thus more efficient than predicting vectorial or tensorial properties directly. In addition, directly outputting polarization or Born charges, as is done in many previous works, fails to enforce conservation laws and physical sum rules. Third, the $O(3)$ symmetry group equivariance of our model yields high accuracy, data efficiency and scalability \cite{loose2023JPCB, fu2023arxiv, maxson2024arxiv}. This is particularly valuable when learning polarization and Born charges, which require as training data computationally expensive DFT calculations in the presence of electric fields. Furthermore, MLMD can be conducted at arbitrary electric field using the model trained to predict correct linear response under zero electric field conditions, eliminating the necessity for training separate models at different electric field magnitudes. This offers advantages in terms of computational cost and memory. Finally, our formalism can be implemented also on ML architectures based on invariant representations, such as Gaussian process regression frameworks \cite{vandermause2020npj}. Equivariance of the model is a useful benefit but not a necessary architectural element, as it would be in the case of direct prediction of vector and tensor quantities.

\vspace{0.3cm}
\noindent \textbf{Simulations with external electric field}\\
In the case of a constant electric field, the electric enthalpy is obtained as in equation \eqref{eq:enthalpy}. The forces are calculated as
\begin{equation}
\mathbf F_{i} = \mathbf F_{i}^0 + e \mathbf Z_i \cdot \mathbf  E, \label{eq:force_efield}
\end{equation}
where $\mathbf F_{i}^0$ is the force on atom $i$ in the absence of the electric field. The polarization is given by
\begin{equation}
    \mathbf P = \mathbf P^0 + \bm\alpha \cdot \mathbf E, \label{eq:polarization_efield}
\end{equation}
where $\mathbf P^0$ is the polarization in the absence of the electric field. In equations \eqref{eq:force_efield}  and \eqref{eq:polarization_efield},  $\mathbf F_i$ and $\mathbf P$ obtained in the presence of an electric field are calculated using quantities determined in the limit of zero electric field, namely $\mathbf F^0$, $\mathbf P^0$, $\mathbf Z_i$ and $\bm \alpha$. This stems from the linearity of the electric enthalpy functional with respect to small electric fields in the modern theory of polarization. The inclusion of the electric-field contributions in equations \eqref{eq:enthalpy}, \eqref{eq:force_efield}, and 
\eqref{eq:polarization_efield} is implemented in our \textsc{lammps} interface, which works with for both time-dependent and space-dependent electric fields. In particular, to calculate the ferroelectric hysteresis of \ce{BaTiO3} at 0~K through structural relaxations, we use the following electric field along $\mathbf z$ 
\begin{equation}
E(t) = E_\text{max} \cos\left(2\pi \left\lfloor\frac{t}{t_0}\right\rfloor\frac{t_0}{\tau} \right), \label{eq:efield_t}
\end{equation}
where $E_\text{max} = 36~\text{MV}\cdot{\text{cm}}^{-1}$, $t_0 = 50$ steps is the interval over which the electric field is kept constant to allow for the convergence of the structural relaxation, and $\tau = 20000$ steps is the period of the cosine. For finite-temperature MD, the electric field is a cosine function, namely
\begin{equation}
E(t) = E_\text{max} \cos\left(2\pi \frac{t}{\tau}\right), \label{eq:efield_t_MD}
\end{equation}
where $E_\text{max} = 36~\text{MV}\cdot{\text{cm}}^{-1}$, and $\tau = 200~\text{ps}$. The time step of the MLMD simulation is set to 2~fs. For analyzing the polarization dynamics of \ce{BaTiO3}, as Born charges are essentially constant throughout the MD, we  assign a dipole for each unit cell $u$ using the formula
\begin{equation}
\mathbf P^{(u)} = \frac 12 \sum_{\ce{O} \in u} \mathbf Z^*_{\ce{O}} \cdot \Delta \mathbf r^{(u)}_{\ce{O}-\ce{Ti}} + \frac 18 \sum_{\ce{Ba} \in u} \mathbf Z^*_{\ce{Ba}} \cdot \Delta \mathbf r^{(u)}_{\ce{Ba}-\ce{Ti}}, \label{eq:dipoles_unitcell}
\end{equation}
where $\Delta \mathbf r^{(u)}_{\ce{O}-\ce{Ti}}$ and $\Delta \mathbf r^{(u)}_{\ce{Ba}-\ce{Ti}}$ denote the coordinates of O and Ba atoms relative to the Ti atom in the unit cell $u$, respectively, and $\mathbf Z^*_{\ce{O}}$ and $\mathbf Z^*_{\ce{Ba}}$ are the respective Born charge tensors. In equation \eqref{eq:dipoles_unitcell}, we consider local dipoles of 5-atom unit cells with one Ti atom at the center, eight Ba atoms at the corners, and six O atoms at the center of all faces. \\

\vspace{0.3cm}
\noindent \textbf{Physical symmetries and conservation laws}\\
We demonstrate that in our model physical symmetries and conservation laws stem from the enforcement of the translation invariance of the generalized potential energy combined with the exact derivative relations between physical quantities [equations \eqref{eq:pol_E}-\eqref{eq:polarizability}]. In our model, the generalized potential energy is translation invariant as it depends on interatomic displacements \cite{musaelian2023NC}. Then, calculating forces as gradient of the generalized potential energy ensures momentum conservation. Indeed, the translation invariance of the generalized potential energy can be expressed as
\begin{equation}
\nabla_{\mathbf c} U(r_{i\mu} + c_{\mu})  = -\sum_{i=1}^N \mathbf F_{i} = 0, \label{eq:gradient_E}
\end{equation}
where $\mathbf c$ is an arbitrary displacement vector, and where we used the chain rule to differentiate with respect to atomic positions. Forces summing up to zero ensure momentum conservation. In addition, in our model, polarization is translation invariant as it is calculated as a gradient of the generalized potential energy with respect to the electric field. This ensures the charge neutrality condition for Born charges. Indeed, following the same reasoning as in equation \eqref{eq:gradient_E}, the translation invariance of polarization can be expressed as
\begin{equation}
\nabla_{\mathbf c} \mathbf P(r_{i\mu} + c_{\mu})  = \sum_{i=1}^N \mathbf Z_{i} = 0, \label{eq:gradient_P}
 \end{equation}
which corresponds to the acoustic sum rule for Born charges. Next, the determination of forces as gradients of the generalized potential energy with respect to atomic positions guarantees the electric enthalpy conservation in MLMD. Indeed, 
\begin{equation}
\frac{\partial U}{\partial t} = - \sum_{i=1}^N \mathbf F_i \cdot \dot{\mathbf r}_i = - \frac{\partial K}{\partial t} ,  \label{eq:dE_dt}
\end{equation}
where $K$ is the kinetic energy. Finally, any cyclic adiabatic evolution involving changes in the electric field yields zero electric work due to the conservative nature of polarization, namely:
\begin{equation}
\oint \mathbf P \cdot d\mathbf E= - \oint \nabla_{\mathbf E} U \cdot d\mathbf E =  0, \label{eq:P_dEfield}
\end{equation}
where we used the fact that in our model polarization is determined  as gradient of the electric enthalpy with respect to the electric field. The enforcement of equation \eqref{eq:P_dEfield} implies that the ferroelectric hysteresis loop is exactly symmetric with respect to reversing the direction of the electric field, as obtained in Fig.\ \ref{fig:results_BaTiO3}a-b.

Several works \cite{christensen2019JCP, gastegger2021CS, fang2024arxiv} calculate dipole moments of isolated molecules as gradients of the energy with respect to the electric field, which enters in the input representation of model, conceptually similarly to our work. 
At variance, our approach applies to both molecular and extended systems by ensuring that the electric field enters as an input to the model to predict the electric enthalpy (or other generalized potentials), and the training is carried out over electric enthalpy, forces and dielectric response properties such as polarization, Born charges and polarizability. Enforcement of the acoustic sum rule for Born charges is enforced in several prior works where Born charges are calculated as derivatives of polarization with respect to atomic displacement, \cite{zhang2022JCP, joll2024arxiv, shimizu2023STAM, fang2024arxiv, schmiedmayer2024arxiv}. In Refs.\ \cite{zhang2022JCP,shimizu2023STAM}, where MLMD under electric fields are conducted, the electric enthalpy is conserved due to the fact that forces are calculated as gradient of the electric enthalpy. However, the methods in Refs.\ \cite{zhang2022JCP,joll2024arxiv,shimizu2023STAM,schmiedmayer2024arxiv} require the use of two separate models for determining the force field and the dielectric properties, and therefore incur the corresponding computational overhead. At variance, in our work, polarization and Born charges are predicted together with energy and forces within a unified model that preserves the conservative nature of polarization, the acoustic sum rule, and the electric enthalpy conservation. These properties are either not enforced or not considered in other previous works \cite{veit2020JCP,krishnamoorthy2021PRL,staackeMLST2022,gigli2022npj,schutt2021ICML,kapil2023FD,schienbein2023JCTC,shao2023NC,sommers2020PCCP, choudhary2020npj,wilkins2019PNAS,takahashi2020PRM,berger2024PRM}. 

\vspace{0.3cm}
\noindent \textbf{Training data generation}\\
First, we collect a set of uncorrelated frames. For $\alpha$-\ce{SiO2}, this is achieved through classical MD simulations using the Vashishta potential \cite{broughton1997PRB} in the NVT ensemble. These simulations are performed at both 300~K and 600~K, each for 100 ps. We take 100 uncorrelated snapshots from each MD at intervals of 1 ps, yielding a total of 200 frames. For \ce{BaTiO3}, we collect a total of 75 frames using active learning dynamics using the FLARE code  \cite{vandermause2020npj}, with temperature ranging from 300~K to 400~K. In particular, 60 frames are collected through active learning MD starting from a pristine structure, and additional 15 frames are collected through active learning MD starting from a domain wall structure. Next, for each frame, we perform DFT calculations to determine energy, forces, and polarization in the absence of electric field. To calculate Born charges and polarizability, we perform DFT calculations in the presence of small uniform electric fields, and use finite differences involving forces and polarization, respectively. In particular, the Born charges are calculated through the following expression
\begin{equation}
    Z^*_{i \mu \nu} = \left.\frac{1}{e} \frac{\partial F_{i\nu}}{\partial E_\mu}\right|_{E_\mu = 0},\label{eq:born_charge_2}
\end{equation}
which derives from combining equation \eqref{eq:pol_E} and the definition of atomic forces $F_{i \nu} = - \partial U/\partial r_{i \nu}$. The polarizability is determined as in equation \eqref{eq:polarizability}. The DFT calculations are performed using a plane-wave density functional approach as implemented in the \textsc{Quantum Espresso} suite \cite{giannozzi2009JPCM}. A small electric field of $0.36~\text{MV}\cdot\text{cm}^{-1}$ is used, in order to ensure the linear regime of polarization with respect to electric field. Additional computational details are provided in the Supplementary Information (SI).

\vspace{0.3cm}
\noindent \textbf{Vibrational and dielectric properties from MD}\\
We discuss the determination of vibrational and dielectric properties from MLMD. The infrared spectrum and the frequency-dependent dielectric constant can be determined from a MD simulation in the absence of the electric field at a given temperature \cite{neumann1983MP,neumann1983CPL}. In particular, the infrared absorption spectrum is calculated as \cite{marsalek2017PRL} 
\begin{equation}
    I(\omega) \propto \omega^2 \text{Re}\left[\int_{0}^{+\infty} dt \, e^{-i\omega t} \braket{\mathbf P^0(t) \cdot \mathbf 
 P^0(0)}\right], \label{eq:infrared}
\end{equation}
where $\omega$ is the frequency, $t$ the time, $\mathbf P^0$ the polarization in the absence of an electric field, and $\braket{\mathbf P^0(t) \cdot \mathbf P^0(0)}$ the average of the autocorrelation function of polarization. The polarizability can be used to determine the high-frequency dielectric constant through the following expression
\begin{equation}
\varepsilon^\infty_{\mu\nu} = 1 + \frac{4\pi}{\Omega} \braket{\alpha_{\mu\nu}}, \label{eq:epsiloninf_MD}
\end{equation}
where $\Omega$ is the volume, and $\braket{\alpha_{\mu\nu}}$ the average polarizability. The static dielectric constant can then be determined by adding an ionic contribution related to the polarization during the MD through the fluctuation-dissipation theorem, namely 
\begin{equation}
\varepsilon^0_{\mu\nu} = \varepsilon^\infty_{\mu\nu} + \frac{4\pi}{\Omega} \frac{\text{cov}(P^0_\mu,P^0_{\nu})}{k_\text{B}T}, \label{eq:epsilon0_MD}
\end{equation}
where $k_\text{B}$ is the Boltzmann constant, $T$ the temperature, and $\text{cov}(P^0_\mu,P^0_\nu)$ the covariance of $P^0_\mu$ and $P^0_\nu$. Then, one can determine the frequency-dependent dielectric constant using the autocorrelation function of polarization as follows:
\begin{align}
    \varepsilon_{\mu\nu}(\omega) & = 1 + (\varepsilon^0_{\mu\nu} -1) \cdot  \nonumber \\
    & \quad \cdot \Bigg[ 1 - i\omega \int_0^{+\infty} dt\, e^{-i\omega t} \frac{\braket{P^0_\mu(t)  P^0_\nu(0)}}{\text{cov}(P^0_\mu,P^0_\nu)} \Bigg].\label{eq:eps_omega}
\end{align}
In Fig.\ \ref{fig:results_SiO2}d, the dielectric constant is calculated as
\begin{equation}
\varepsilon_{\mu\nu} = 1 + \frac{4\pi}{\Omega} \frac{P_{\mu}(\mathbf R, \mathbf E) - P_{\mu}(\mathbf R^0, 0)}{E_\nu}, \label{eq:epsilon_P}
\end{equation}
where $P_{\mu}(\mathbf R,\mathbf E)$ is the polarization obtained in the presence of an electric field $\mathbf E$ for a given structure $\mathbf  R$, and $P_{\mu}(\mathbf R^0,0)$ is the polarization obtained in the absence of the field for the initial structure $\mathbf R^0$. The structure $\mathbf R^0$ is found by performing DFT relaxations in the absence of the electric field. In Fig.\ \ref{fig:results_SiO2}d, the structure $\mathbf R$ is obtained relaxing the system in the presence of the field $\mathbf E$ along $\mathbf z$. Equations \eqref{eq:epsiloninf_MD}-\eqref{eq:epsilon_P} are written in atomic units. In eV units, $4\pi$ is replaced with $1/\epsilon_0$, where $\epsilon_0$ is the vacuum permittivity. 

\vspace{0.3cm}
\noindent \textbf{\large Acknowledgements} \\ 
We thank L.\ Giacomazzi for insight on DFPT calculations, and N. Rivano, Z.\ Goodwin,  B.\ Duschatko, and S.\ Kavanagh for useful discussions. This work was supported primarily by the NSF through the Harvard University Materials Research Science and Engineering Center Grant No.\ DMR-2011754, US Department of Energy, Office of Basic Energy Sciences Award No.\ DE-SC0022199 as well as by the Camille and Henry Dreyfus Foundation Grant No. ML-22-075, the Department of Navy award N00014-20-1-2418 issued by the Office of Naval Research and Robert Bosch LLC. Computational resources were provided by the Harvard University FAS Division of Science Research Computing Group. S.F.\ was supported by the Swiss National Science Foundation through the Postdoc mobility fellowship  under grant number P500PT\_214445. C.J.O.\ was supported by the National Science Foundation Graduate Research Fellowship Program under Grant No.\ (DGE1745303). A.J. was supported by Aker Scholarship. Computational resources were provided by the FAS Division of Science Research Computing Group at Harvard University. Additional resources include the National Energy Research Scientific Computing Center (NERSC), a DOE Office of Science User Facility supported by the Office of Science of the U.S.\ Department of Energy under Contract No.\ DE-AC02-05CH11231 using NERSC award BES-ERCAP0024206.

\vspace{0.3cm} \noindent {\large \textbf{Author contributions}} \vspace{0.1cm}\\
S.F., A.C., C.W.T., A.M., and B.K.\ jointly conceived the architecture of the ML model. S.F.\ did the DFT and DFPT calculations, contributed to the implementation of the ML architecture, trained the ML models, conceived the applications, analyzed the results, prepared the figures, and wrote the first version of the manuscript. A.J.\ implemented the \textsc{lammps} interface, and designed and ran the MD and hysteresis simulations. C.W.T. contributed to the implementation of the ML architecture, and to the analysis of the results. A.M.\ did the majority of the implementation of the ML architecture. C.J.O.\ contributed to the optimization of the ML models. B.K.\ supervised and guided the project from conception to analysis of results. All authors contributed to the manuscript.

\vspace{0.3cm} \noindent {\large \textbf{Competing interests}} \vspace{0.1cm}\\
The authors declare no competing interests.

\newpage
\onecolumngrid
\hspace*{-2.4cm}
\includegraphics[page=1, scale=1.0]{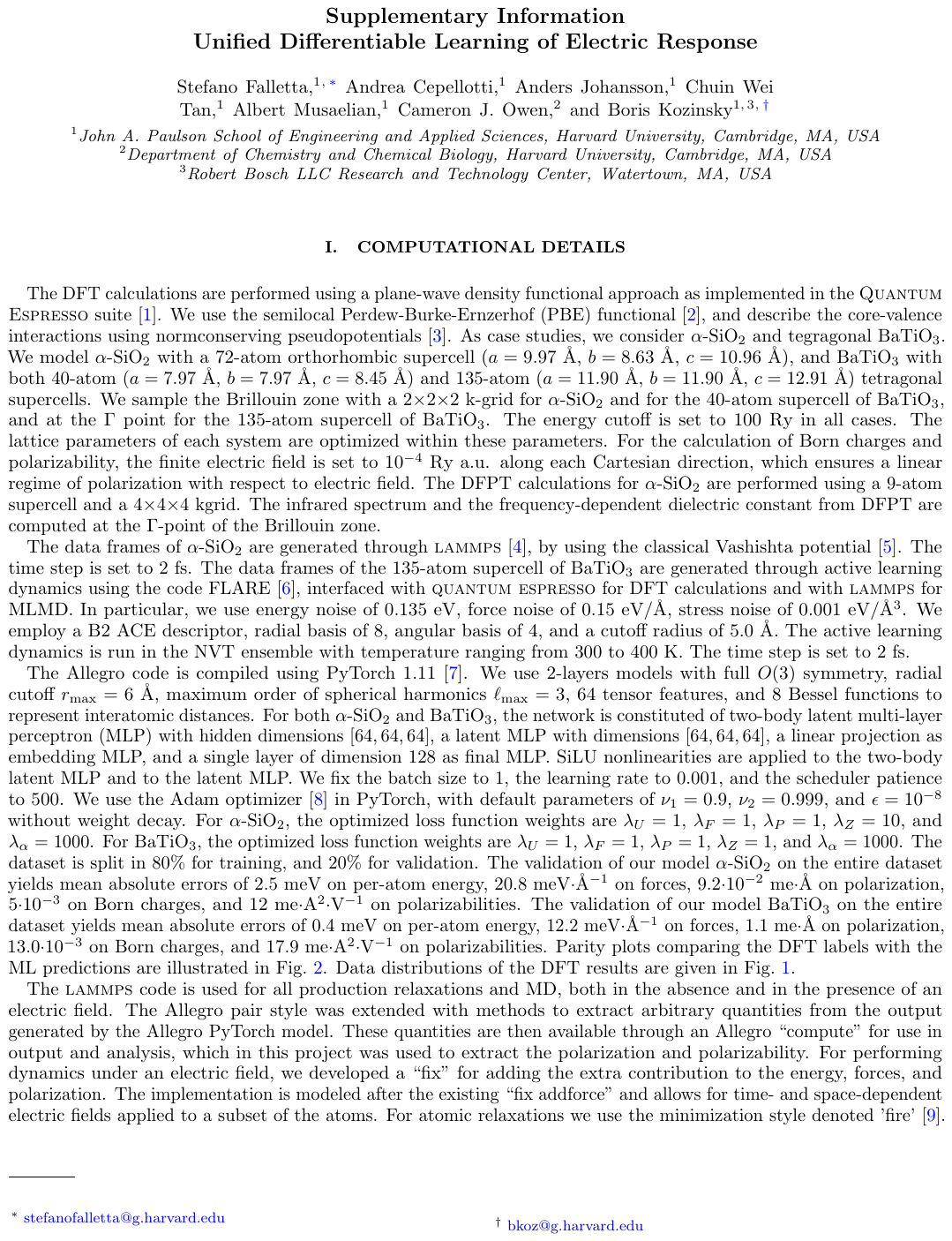}
\hspace*{-2.4cm}
\includegraphics[page=2, scale=1.0]{SI.pdf}
\hspace*{-2.4cm}
\includegraphics[page=3, scale=1.0]{SI.pdf}

\end{document}